\documentclass[twocolumn,twoside]{IEEEtran}
\usepackage{mathpple}
\usepackage{times}
\usepackage{amsmath}  
\usepackage{amssymb}  
\usepackage{mathrsfs} 
\usepackage{theorem}  
\usepackage{comment}  
\usepackage{upref}
\usepackage{amsfonts}
\usepackage{verbatim}
\usepackage[dvipsnames,usenames]{color}
\usepackage{enumerate}
\usepackage{graphicx}
\usepackage{subfigure}
\usepackage{latexsym}
\usepackage{mathtools} 
\usepackage{array, tabularx} 
\usepackage[shortlabels]{enumitem}
\usepackage{url}
\usepackage{amsfonts} 

\usepackage{bbm}
\usepackage{times,color}
\usepackage{enumitem}
\usepackage{float}
\usepackage{listings}
\usepackage{tikz}
\usetikzlibrary{matrix}
\usepackage{multicol}
\usepackage{psfrag}
\usepackage{pgfplots}
\usepackage{hhline}
\usepackage{xcolor,colortbl}
\usepackage{algorithm,algpseudocode,refcount} 

\renewcommand{\algorithmicrequire}{\textbf{Input:}}
\renewcommand{\algorithmicensure}{\textbf{Output:}}
\algnewcommand\algorithmiccase{\textbf{case}}
\algdef{SE}[CASE]{Case}{EndCase}[1]{\algorithmiccase\ #1}{\algorithmicend\ \algorithmiccase}%
\algtext*{EndCase}%
\makeatletter
\newcommand{\removelatexerror}{\let\@latex@error\@gobble}

\newcommand{\limup}[1]{\lim_{#1\rightarrow\infty}}

\newcommand{\limsupup}[1]{\limsup_{#1\rightarrow\infty}}

\newcommand{\1}{\mathbbm{1}}
\newcommand{\bfone}{\mathbf{1}}
\newcommand{\0}{\mathbf{0}}

\newcommand{\be}[1]{\begin{equation}\label{#1}}
\newcommand{\ee}{\end{equation}}

\newcommand{\bc}{\begin{center}}
	\newcommand{\ec}{\end{center}}

\newcommand{\floorenv}[1]{\left\lfloor #1 \right\rfloor}

\newcommand{\ceilenv}[1]{\left\lceil #1 \right\rceil}

\makeatletter
\renewcommand{\thefigure}{{\@arabic\c@figure}}
\renewcommand{\fnum@figure}{{\bf Figure\,\thefigure}}
\makeatother

\newcommand{\cA}{{\cal A}}

\newcommand{\cC}{{\cal C}}
\newcommand{\cD}{{\cal D}}

\newcommand{\cO}{{\cal O}}

\newcommand{\cS}{{\cal S}}

\newcommand{\cZ}{{\cal Z}}

\newcommand{\bfe}{{\boldsymbol e}}

\newcommand{\bfu}{{\boldsymbol u}}
\newcommand{\bfv}{{\boldsymbol v}}

\newcommand{\bfx}{{\boldsymbol x}}
\newcommand{\bfy}{{\boldsymbol y}}
\newcommand{\bfz}{{\boldsymbol z}}

\renewcommand{\le}{\leqslant}
\renewcommand{\leq}{\leqslant}
\renewcommand{\ge}{\geqslant}
\renewcommand{\geq}{\geqslant}

\newcommand{\Cref}[1]{Co\-rol\-la\-ry\,\ref{#1}}

\theoremstyle{plain} \theorembodyfont{\normalfont\slshape}

\newtheorem{thm}{Theorem$\!$}
\newenvironment{theorem}{\begin{thm}\hspace*{-1ex}{\bf.}}{\end{thm}}

\newtheorem{prop}[thm]{Proposition$\!$}

\newtheorem{lem}[thm]{Lemma$\!$}
\newenvironment{lemma}{\begin{lem}\hspace*{-1ex}{\bf.}}{\end{lem}}

\newtheorem{cor}[thm]{Corollary$\!$}
\newenvironment{corollary}{\begin{cor}\hspace*{-1ex}{\bf.}}{\end{cor}}

\newtheorem{cons}[thm]{Construction$\!$}

\newtheorem{defi}[thm]{Definition$\!$}
\newenvironment{definition}{\begin{defi}\hspace*{-1ex}{\bf.}}{\end{defi}}

\newtheorem{cl}[thm]{Claim}
\newenvironment{claim}{\begin{cl}\hspace*{-1ex}{\bf .}}{\end{cl}}
\theorembodyfont{\normalfont}

\newtheorem{exam}{Example$\!$}
\newenvironment{example}{\begin{exam}\hspace*{-1ex}{\bf .}}{\end{exam}}

\newtheorem{remrk}{Remark$\!$}
\newenvironment{remark}{\begin{remrk}\hspace*{-1ex}{\bf .}}{\end{remrk}}

\definecolor{Codecolor}{named}{White}  

\newcommand{\Copen}{\mbox{\{\kern-5.50pt\{}}
\newcommand{\Cclose}{\mbox{\}\kern-5.50pt\}}}
\newcommand{\Cslash}{\mbox{$\backslash\kern-6.02pt\backslash$}}

\newcommand{\red}[1]{\text{red}({#1})}
\newcommand{\ered}[1]{\emph{\text{red}}({#1})}

\begin{document}
	\title{The Zero Cubes Free and Cubes Unique Multidimensional Constraints}
	
	\author{\large Sagi~Marcovich,~\IEEEmembership{Student Member,~IEEE} and Eitan~Yaakobi,~\IEEEmembership{Senior Member,~IEEE} 
		\thanks{S. Marcovich and E. Yaakobi are with the Department of Computer Science, Technion --- Israel Institute of Technology, Haifa 3200003, Israel (e-mail: \texttt{\{sagimar,yaakobi\}@cs.technion.ac.il}).}
	}

	\maketitle
	\begin{abstract}
		This paper studies two families of constraints for two-dimensional and multidimensional arrays. The first family requires that a multidimensional array will not contain a cube of zeros of some fixed size and the second constraint imposes that there will not be two identical cubes of a given size in the array. These constraints are natural extensions of their one-dimensional counterpart that have been rigorously studied recently. For both of these constraint we present conditions of the size of the cube for which the asymptotic rate of the set of valid arrays approaches 1 as well as conditions for the redundancy to be at most a single symbol. For the first family we present an efficient encoding algorithm that uses a single symbol to encode arbitrary information into a valid array and for the second family we present a similar encoder for the two-dimensional case. The results in the paper are also extended to similar constraints where the sub-array is not necessarily a cube, but a box of arbitrary dimensions and only its volume is bounded. 
	\end{abstract}

	\section{Introduction} \label{sec:intro}
	Coding for two-dimensional and multidimensional arrays is a topic which attracted significant attention in the last three decades due to its various applications in different areas. 
This includes optical storage such as page-oriented optical memories~\cite{Heanue749,NeMc94} and holographic storage~\cite{Gu92}. Other applications in robotics are robot localization~\cite{S01}, camera localization~\cite{S12}, projected touchscreens~\cite{DC14}, just to name a few, and there are several more in structured light; see e.g.~\cite{Hs01,MoOzCo98,SaFePr10, PaSaCo05}. Examples of coding schemes for these applications include error-correction codes~\cite{EY09}, constrained codes~\cite{TR10}, pseudo random arrays and perfect maps~\cite{MaSl76, Et88}, codes for self locating patterns~\cite{BrEtGi2012}, and more. 

This paper takes one more step in advancing the theory of coding for multidimensional and studies two special constraint families for two-dimensional and multidimensional arrays. In the first constraint, it is said that an array $ W \in \Sigma_q^{[n]^d} $ is \emph{zero $ L $-cubes free} if it does not contain any zero cube of volume $L^d$. In the second constraint, we say that $W$ is \emph{$ L $-cubes unique} if it does not contain any two identical cubes of volume $L^d$.
Only little is known on these families of codes and the goal of this paper is to rigorously study them for all values of $L$ and $d$ and in particular for $ d = 2 $, as well as to construct efficient encoding and decoding algorithms for these constraints. 

The zero $L$-cubes free constraint was studied for the one-dimensional case in~\cite{LevYaa18}. It was shown that if $ L = \log_q (n) - f(n) $ where $ f(n) $ satisfies that $n-2(\log_q (n) - f(n)) = \Theta(n)$ then the redundancy of the sequences that satisfy the constraint is $ \Theta(q^{f(n)}) $. An encoding scheme for the binary case that uses a single redundancy bit and avoids zero-runs of length $L = \lceil \log(n) \rceil + 1$ was also proposed.  The $L$-cubes unique constraint was studied in~\cite{EliGabMedYaa19, EliGabMedYaa19IEEE} and it was shown that for $ L $ that satisfies $ L^d~=~\floorenv{ ad \log_q(n)} $ with $ a > 1 $, the asymptotic rate of these arrays approaches 1. For the one-dimensional case, two encoding schemes were proposed. The first one uses a single redundancy symbol and supports $ L = 2\lceil \log (n)\rceil + 2$, while the second works for substrings of length $L=\lceil a \log (n) \rceil$ where $ 1 < a \le 2 $ and its asymptotic rate approaches 1. 

In this paper it is shown that for the zero $L$-cubes free constraint, if $L = \omega(1)$ then the asymptotic rate of the set of arrays that satisfy the constraint approaches 1 and for $ L \geq \sqrt[d]{d\log_q(n)+\log_q(\frac{q}{q-1})}$, its redundancy is at most a single symbol. Then, an efficient algorithm for encoding $L$-cubes free arrays that uses a single redundancy symbol is presented for $ L =\ceilenv{\sqrt[d]{\ceilenv{d\log_q(n)}+1}}$. Note that the difference between these two values of $L$ is at most $ 1 + \sqrt[d]{2} $. 
Moreover, for the two-dimensional case it is proven that if $ n - 2L = \Theta(n) $ then the redundancy of the arrays that satisfy the constraint is $ \Theta\left(n^2/q^{L^2}\right) $.
For the $L$-cubes unique constraint, it is shown that if $L \geq \sqrt[d]{2d \log_q (n) + \log_q(\frac{q}{q-1})}$, then the redundancy of this set of arrays is at most a single symbol. For the binary two-dimensional case, an encoding algorithm that uses a single binary bit is proposed which supports $ L = 2 \ceilenv{\sqrt{\ceilenv{3 \log (n)} +2}}$. 

Later, the paper tackles the extensions of the aforementioned constraints for the case where only the volume of the sub-arrays is constrained so their shape is not necessarily a cube, but a box with a coordinates set $ [x_0] \times \cdots \times [x_{d-1}] $ and $\prod_{i=1}^dx_i\geq V$ and $V$ is the constraint's parameter. Namely, it is said that an array $ W \in \Sigma_q^{[n]^d} $ is \emph{zero $ V $-boxes free} if it does not contain any zero box with volume at least $ V $. Similarly, $W$ is called \emph{$ V $-boxes unique} if it does not contain any two identical boxes with volume at least $ V $. As far as we know, these constraints were not studied before. 

In order to study these two constraints, we first bound for any $ d,V $ the number of minimal $ d $-dimensional $ V $-boxes, that is, boxes with volume at least $ V $ that are not contained in any other box of such volume. It is shown that for fixed $ d $, there are $ \Theta(V^{\frac{d-1}{d}}) $ minimal $ V $-boxes. Then, for the zero $ V $-boxes free constraint, it is shown that if $ V \ge d \log_q(n) + \frac{d-1}{d}\log_q(\log_q(n))+ \cO(1), $ the redundancy of the arrays that satisfy the constraint is at most a single symbol. Then, an efficient encoding algorithm for zero $ V $-boxes free arrays that uses a single redundancy symbol is presented for a similar value of $ V $. Additionally, the two-dimensional case is examined and it is proven that if $ n-2\sqrt{V} = \Theta(n)$, then the redundancy of the arrays that satisfy the constraint is $ \Theta\left(n^2/q^{V - \log_q(V)}\right) $. As for the $ V $-boxes unique constraint, it is shown that if $ V \ge 2d \log_q(n) + \frac{d-1}{d}\log_q(\log_q(n))+ \cO(1), $ the redundancy of this family of arrays is at most a single symbol. Furthermore, it is proven that for $ V = ad\log_q(n) $ with $ a > 1 $, the asymptotic rate of the arrays that satisfy the constraint approaches~$ 1 $.

The rest of the paper is organized as follows. In Section~\ref{sec:def}, the constraints that will be studied in the paper are formally defined. In Section~\ref{sec:no-zero-squares}, we study the zero cubes free constraint and in Section~\ref{sec:no-identical-squares} we address the cubes unique constraint. Then, in Section~\ref{sec:ext-v} we study the extensions of these constraints to the zero $ V $-boxes free and $ V $-boxes unique constraints. Section~\ref{sec:red} refines the study of the zero-free constraints for the two-dimensional case and provides an exact asymptotic analysis of their redundancy. Lastly, Section~\ref{sec:concl} concludes the paper.

	\section{Definitions and Preliminaries}\label{sec:def}
	In this section we formally define the notations and constraints studied in this paper. For integers $ i, j \in \mathbb{N} $ such that $ i \le j $ we denote by $ [i,j] $ the set $ \{i, i+1, \dots, j-1, j\} $. We notate by $ [i] $ a shorthand for $  [0,i-1] $. For a set $ A $, let $ |A| $ denote the number of elements in $A$. Let $ \Sigma_q $ denote a finite alphabet of size $ |\Sigma_q| = q $. When $ q = 2 $, we omit the subscript $ q $ from this and from similar notations. 

Let $ d \in \mathbb{N} $ be an integer, let $ \mathbb{N}^d $ be the $ d $-dimensional grid, and let $ \bfv = (v_0, v_1, \dots , v_{d-1}) \in \mathbb{N}^d $ denote a vector of length $ d $. For $ A \subseteq \mathbb{N}^d $, a set of coordinate vectors, we denote by $ \bfv + A $ the set $$ \{ (v_0 + u_0, \dots, v_{d-1} + u_{d-1}) \mid \bfu = (u_0, \dots, u_{d-1}) \in A \},$$ 
and by $ c \cdot A $, where $ c \in \mathbb{N} $, 
the set $$ \{ (c u_0, \dots, c u_{d-1}) \mid \bfu = (u_0, \dots, u_{d-1}) \in A \}.$$
The set  $ \bfv - A $ is defined similarly.
Next, for a set $ A \subseteq \mathbb{N}^d $ we denote by $ \Sigma_q^{A} $ the set of all functions from $ A $ to $ \Sigma_q $. We denote by $ \bigcup_{A \subseteq \mathbb{N}^d} \Sigma_q^{A} $ the set of all \emph{$ d $-dimensional arrays}.
For an integer $n\in\mathbb{N}$, we denote by $[n]^d$ the set  $[n]^d=\otimes_{i=0}^{d-1}[n]$ and say that $ \Sigma_q^{[n]^d} $ is the set of all \emph{$d $-dimensional $ n $-cubes}. Throughout this paper, we sometimes remove the $ d $-dimensional prefix when using those notations when the dimension $d$ is clear from the context. When $ d=2 $, we refer to $d $-dimensional $ n $-cubes as $ n $-squares. 
Additionally, the redundancy of a set $ \cA \subseteq \Sigma_q^{ [n]^d} $ is defined as $ \red{\cA} = n^d - \log_q(|\cA|) $.

Let $ W \in \Sigma_q^{A} $  be an array and $ A' \subseteq A \subseteq  \mathbb{N}^d $ be sets of coordinate vectors. We denote by $ W_{A'} $ the restriction of $ W $ to the coordinates in $ A' $. When $ A' $ contains a single coordinate vector $ A' = \{\bfv\} $ we simplify the representation and write $ W_\bfv $. Next, we define a total order over $ \mathbb{N}^d $. 
\begin{definition}\label{def:nd-order}
	Let $  \bfu = (u_0, \dots , u_{d-1}), \bfv = (v_0,\dots, v_{d-1}) \in \mathbb{N}^d $ be two different coordinate vectors. We say that $ \bfu < \bfv $ if there exists $ 0 \le s \le d-1  $ such that $ u_s < v_s $ and for every $ 0 \le t < s $, $ u_t = v_t $. 
\end{definition}

For a set $A \subseteq \mathbb{N}^d $  and a vector $ \bfv \in A $, the mapping $ B_{A,q}(\bfv) $ returns a $ q $-ary vector of the index representation of $ \bfv $ in $ A $, where the vectors are ordered increasingly according to the total order presented in Definition~\ref{def:nd-order}. Note that the size of the mapping output is  $ \ceilenv{\log_q (|A|)}$. For an integer $ i \in [d] $, let $ \bfe_i \in \Sigma_2^d $ denote the $ i $-th unit vector, i.e., a vector with \emph{one} at its $ i $-th bit and \emph{zeros} elsewhere.
Additionally, we denote the bijection $MD_A:\Sigma_q^{|A|}\rightarrow\Sigma_q^{A}$ which transforms a sequence to its multidimensional representation under the coordinates of $ A $, and its inverse $SD_A:\Sigma_q^{A}\rightarrow\Sigma_q^{|A|}$. $ MD_A $ reorders the symbols using the order of Definition~\ref{def:nd-order} over the coordinates of $ A $, i.e., the $ i $-th symbol of the input sequence will transform to the symbol in the $ i $-th coordinate in $ A $. We will sometimes omit $ A $ from the notations when it is clear from the context.
\begin{example}
	Let $ n = 4, d=2 $, and 
	\[
	X = 	\begin{pmatrix}
		1 & 1 & 0 & 0\\
		1 & 0 & 1 & 1\\
		0 & 0 & 0 & 1\\
		0 & 0 & 0 & 1\\
	\end{pmatrix} \in \Sigma^{[n]^2}.
	\] 
	Then, $ SD(X) = 1100101100010001 $.
\end{example}

Next, the main families of constraints that are studied in the paper are defined.

\begin{definition}\label{def:zero}
Let $ W \in \Sigma_q^{[n]^d} $ be a $ d $-dimensional array. We say that $W $ contains a \textbf{zero $ L $-cube} (or \textbf{zero $ L $-square} for $ d=2 $) at position $ \bfv \in [n-L+1]^d $, if $ W_{\bfv + [L]^d}  = \0$. An array $ W $ satisfies the  \textbf{zero $ L $-cubes free constraint} if it does not contain any zero $ L $-cube.
\end{definition}
Throughout the paper, we sometimes refer to an array that satisfies the constraint in Definition~\ref{def:zero} as a \emph{zero $ L $-cubes free} array.

\begin{example}
	Let $ n = 5, d=2 $, and 
	\[
	Y = 	\begin{pmatrix}
	1 & 1 & 0 & 0 & 1\\
	1 & 0 & 1 & 1 & 1\\
	0 & 0 & 0 & 1 & 0\\
	0 & 0 & 0 & 0 & 1\\
	0 & 1 & 0 & 1 & 0\\
	\end{pmatrix} \in \Sigma^{[n]^2}.
	\] 
	Then, $ Y $ contains two zero $ 2 $-squares, at positions $ (2,0)$ and $ (2,1) $. For $ L > 2 $, $ Y $ contains no zero $ L $-squares and thus $ Y $ satisfies the zero $ L $-squares free constraint.

\end{example}

For positive integers $ n,q,d,L $, we denote by $\cC_{d,q}(n,L)$ the set of all arrays over $  \Sigma_q^{[n]^d} $ that satisfy the zero $ L $-cubes free constraint. The authors of~\cite{LevYaa18} studied the one dimensional variation of this problem and showed that if $ L = \log_q (n) - f(n) $,  where $ f(n) $ is a function that satisfies $n-2(\log_q (n) - f(n)) = \Theta(n)$, then the redundancy of $ \cC_{1,q}(n,f(n)) $ is $ \Theta(q^{f(n)}) $. They also proposed an encoding scheme for the binary case that uses a single redundancy bit and avoids zero-runs of length $L = \lceil \log(n) \rceil + 1$.

In Section~\ref{sec:no-zero-squares}, we analyze the cardinality of $\cC_{d,q}(n,L)$ for any $ d,q $,
and present lower bounds for $ L $ for two cases: 1) the asymptotic rate of $\cC_{d,q}(n,L)$ is $ 1$, and 2) the redundancy of $ \cC_{d,q}(n,L) $ is at most a single symbol. Moreover, we present an algorithm that encodes arrays from $\cC_{d,q}(n,L)$ using a single redundancy symbol, where $ L $ almost achieves the lower bound that we found for this case. In Section~\ref{sec:red} we revisit this constraint for the two-dimensional case and give tight bounds for its redundancy for every $ n,L $.

Next, the second constraint studied in the paper is defined. 
\begin{definition}\label{def:unique}
Let $ W \in \Sigma_q^{[n]^d} $ be a $ d $-dimensional array. We say that $W $ contains two \textbf{identical $ L $-cubes}  (or \textbf{identical $ L $-squares} for $ d=2 $) at positions $ \bfu\neq\bfv \in [n-L+1]^d $, if $ W_{\bfu + [L]^d}  =  W_{\bfv + [L]^d}$. An array $ W $ satisfies the  \textbf{$ L $-cubes unique constraint} if it does not contain any two identical $ L $-cubes.
\end{definition}
Throughout the paper, we sometimes refer to an array that satisfies  the constraint in Definition~\ref{def:unique} as an \emph{$ L $-cubes unique} array.

\begin{example}
	Let $ n = 5, d=2 $, and 
	\[
	Z = 	\begin{pmatrix}
		1 & 1 & 0 & 0 & 1\\
		1 & 0 & 1 & 1 & 1\\
		0 & 0 & 1 & 1 & 0\\
		0 & 0 & 1 & 0 & 1\\
		0 & 1 & 0 & 0 & 1\\
	\end{pmatrix} \in \Sigma^{[n]^2}.
	\] 
	Then, $ Z $ contains two identical $ 3 $-squares at positions $ (0,0)$ and $ (2,2) $. However, $ Z $ contains no identical $ 4 $-squares and thus $ Z $ satisfies the $ 4 $-squares unique constraint.
	
\end{example}

We denote by $\cD_{d,q}(n,L)$ the set of all arrays over $ \Sigma_q^{[n]^d} $ that satisfy the $ L $-cubes unique constraint. In~\cite{EliGabMedYaa19IEEE}, the authors analyzed the cardinality of $\cD_{d,q}(n,L)$ and proved the following theorem. 

\begin{theorem}\cite{EliGabMedYaa19IEEE}\label{th:cubes-unique-cardinality}
	For $ L $ that satisfies $ L^d = \floorenv{ad\log_q(n)} $ with $ a > 1 $, the asymptotic rate of $  \cD_{d,q}(n,L) $ approaches 1. Namely, 	$$\lim_{n\rightarrow\infty}\frac{\log_q(|\cD_{d,q}(n,L)|)}{n} =1.$$
\end{theorem}

Additionally, the authors of~\cite{EliGabMedYaa19, EliGabMedYaa19IEEE} proposed two encoding schemes for the one dimensional case of the set $\cD_{1}(n,L)$, which is also known as the set of \emph{$ L $-substring unique} sequences~\cite{MarYaa19,MarYaa19IEEE}. The first scheme is applied for substrings of length $ L = 2\lceil \log (n)\rceil + 2$ with a single bit of redundancy, and the second one works for substrings of length $L=\lceil a \log (n) \rceil$ for any $ 1 < a \le 2 $ and its asymptotic rate approaches 1. In Section~\ref{sec:no-identical-squares}, we  present for all $ d,q $ a lower bound for $ L $ such that the redundancy of $\cD_{d,q}(n,L)$ is at most $ 1 $. Then, we present an encoding scheme for the binary multidimensional case that uses a single redundancy bit, while the value of $ L $ is far from the lower bound we found only by a factor of $ \sqrt{3} $.

	\section{The Zero Cubes Free Constraint}\label{sec:no-zero-squares}
	In this section we study the zero cubes free constraint. We will show in Theorem~\ref{th:rate-cdqnL} a lower bound on $L$ for which the asymptotic rate of the set $\cC_{d,q}(n,L)$ is 1. Then, in Theorem~\ref{th:red-cdqnL}, we find a lower bound on $L$ which implies that the redundancy of the set  $\cC_{d,q}(n,L)$ is bounded from above by 1. Lastly, we present efficient encoding and decoding algorithms that use a single redundancy symbol to encode arrays that are zero $L$-cube free for $ L =\ceilenv{\sqrt[d]{\ceilenv{d\log_q(n)}+1}}. $

We start this section by showing that if $L$ is not a constant then the asymptotic rate of the set $\cC_{d,q}(n,L)$ is $ 1$. 
\begin{theorem}\label{th:rate-cdqnL}
	Let $ L = f(n) $ be a function of $ n $ that is not constant, i.e., $ L = \omega(1) $. Then, the asymptotic rate of $\cC_{d,q}(n,L)$ is $ 1$. Namely,
	$$\lim_{n\rightarrow\infty}\frac{\log_q(|\cC_{d,q}(n,L)|)}{n} = 1.$$
\end{theorem}
\begin{IEEEproof}
	Let $ A $ be the set of coordinates $$ A = ( L \cdot [1,\ceilenv{\frac{n}{L}}-1])^d \subseteq [n]^d, $$ and let $ \cS $ be the following set of arrays,  $$ \cS = \{ X \in \Sigma_q^{[n]^d} \mid \text{ for every } \bfv \in A , X_{\bfv} = 1  \}. $$
	For every $ X \in \cS $, every $ L $-cube contained in $ X $ contains a coordinate from $ A $, and thus $ X $ is zero $ L $-cubes free and $ \cS \subseteq \cC_{d,q}(n,L) $. The size of $ \cS $ satisfies $ |\cS| \ge q^{n^d - \frac{n^d}{L^d}} $, and therefore it is deduced that
	$$\lim_{n\rightarrow\infty}\frac{\log_q(|\cC_{d,q}(n,L)|)}{n} \ge1 - \frac{1}{L^d},$$
	which approaches $ 1 $ for $ L=f(n) $ that is not a constant.
\end{IEEEproof}

Next, we utilize the union bound to reach the following upper bound on the redundancy of the set $\cC_{d,q}(n,L)$.
\begin{theorem}\label{th:red-cdqnL}
	For an integer $ L \geq \sqrt[d]{d\log_q(n)+\log_q(\frac{q}{q-1})} $, it holds that $ |\cC_{d,q}(n,L)| \ge q^{n^d-1} $. That is, $ \emph{\textmd{red}} (\cC_{d,q}(n,L) )\leq 1$. 
\end{theorem} 
\begin{IEEEproof}
	Let $ W \in \Sigma_q^{[n]^d} $ be an array. If $ W $ is not zero $ L $-cubes free, then it contains at least one zero $ L $-cube. Therefore, according to the union bound, the number of arrays over $ \Sigma_q^{[n]^d} $  that are not zero $ L $-cubes free can be bounded from above by
	\begin{align*}
	n^{d} q^{n^d-L^d} = q^{n^d} \cdot \frac{n^{d}}{q^{L^d}} & \leq  q^{n^d} \cdot \frac{n^{d}}{q^{d \log_q(n)} \cdot q^{ \log_q(\frac{q}{q-1})} } \\&= (q-1)q^{n^d-1},
	\end{align*}
	where the inequality follows from the lower bound on $L$ stated in the theorem. This implies that $$ |\cC_{d,q}(n,L)| \ge q^{n^d} - (q-1)q^{n^d-1} = q^{n^d-1}. $$ 
\end{IEEEproof}

Our next goal in the paper is to provide an algorithm that encodes $ d $-dimensional arrays over $ \Sigma_q^{[n]^d} $ which satisfy the zero $ L $-cubes constraint for $$ L =\ceilenv{\sqrt[d]{\ceilenv{d\log_q(n)}+1}}. $$
Note that the difference between this value of $ L $ and the lower bound derived in Theorem~\ref{th:red-cdqnL} is at most $ 1 + \sqrt[d]{2} $. The algorithm uses a single redundancy symbol and its encoding and decoding time complexities is $ O(d  n^d\log (n) ) $. 

Algorithm~\ref{alg:zero-cubes} receives a $ d $-dimensional array $W\in\Sigma_q^{[n]^d\setminus\{ (n-1) \cdot \bfone \}}$ with a single symbol missing at its corner, and outputs a cube $X \in \cC_{d,q}(n,L)$. First, we initialize $ X $ with $ W $ and set  1 at the missing entry to mark the start of the algorithm. Then, we scan over all $ L $-cubes in $ X $ from start to end and look for a zero $ L $-cube. When such a cube is found, it is replaced with the non-zero cube at the position $(n-L)\cdot \bfone $ which will be referred as the \emph{lookup-cube}. The lookup-cube is then filled with an encoding of the position of the zero cube that was found and at least one more additional zero symbol to mark the occurrence of the zero cube to the decoding process. In the case which the found cube and the lookup-cube intersect, we backup only the non-intersecting part of the lookup-cube, since we know the rest of it is zero.

\begin{algorithm}
\caption{Zero $ L $-Cubes Free Encoding}\label{alg:zero-cubes}
\algorithmicrequire{ A $ d $-dimensional array $W\in\Sigma_q^{[n]^d\setminus\{ (n-1) \cdot \bfone \}}$} \\
\algorithmicensure { A $ d $-dimensional array $X \in \cC_{d,q}(n,L)$}
\begin{algorithmic}[1]
	
	\State{Set an array $X \in \Sigma^{[n]^d}$ with $X_{[n]^d\setminus\{ (n-1) \cdot \bfone \}} = W $ and $ X_{(n-1) \cdot \bfone}=1$}\label{step:init-xy}
	\For{every $\bfv  \in [n-L+1]^d $ (iterate in an increasing order)} \label{step:zc-loop}

	\If{$X_{\bfv+[L]^d} = \0$}\label{step:test-0}
	\If{ $ A = (\bfv+[L]^d) \cap [n-L,n-1]^d  = \emptyset $ }
	\State{Set $X_{\bfv+[L]^d} = X_{[n-L,n-1]^d}$}\label{step:nzc-replace1}
	\Else 
	\State{Set $ \bfy = SD(X_{([n-L,n-1]^d \setminus A)}) $}	
	\State{Set $X_{\bfv+[L]^d \setminus A} = MD(\bfy)$}\label{step:nzc-replace2}		
	\EndIf 
	\State{Set  \vspace{-2ex} $$X_{[n-L,n-1]^d} = MD(B_{[n]^d,q}(\bfv+\bfe_{d}) \circ 0^{L^d - \ceilenv{d \log_q (n)} })\vspace{-3ex}$$}\label{step:encode}
	\EndIf
	\EndFor

\end{algorithmic}
\end{algorithm}

The correctness of Algorithm~\ref{alg:zero-cubes} is proved in the next lemma.
\begin{lemma}
	Algorithm~\ref{alg:zero-cubes} successfully outputs an array which is zero $ L $-cubes free.
\end{lemma}

\begin{IEEEproof}
	First, notice that throughout the algorithm assignments are correctly defined and the size of $X$ remains $n^d$. 	
	Next, we observe that at each iteration of the for loop, if the condition at Step~\ref{step:test-0} is satisfied, the found zero $ L $-cube is replaced with a non-zero cube, while new zero $ L $-cubes can not be created. The lookup-cube $ X_{[n-L,n-1]^d} $ is initialized as non-zero at Step~\ref{step:init-xy}, and being kept non-zero after every iteration since $ B_{[n]^d,q}(\bfv+\bfe_d) > \0 $ for every $\bfv \in [n-L+1]^d$.
	Thus, it is ensured that at Step~\ref{step:nzc-replace1} we replace a zero cube with a non-zero cube. This also holds for Step~\ref{step:nzc-replace2} in which the found zero cube intersects with the lookup-cube, since $X_{\bfv+[L]^d \setminus A}$ is filled with the non-zero data part of the lookup-cube. Therefore, since we iterate over all the $ L $-cubes in $ X $ at Step~\ref{step:zc-loop}, when the algorithm ends there are no zero $ L $-cubes left. Lastly, note that since $L^d > \ceilenv{d \log_q (n)}$, Step~\ref{step:encode} is successful and after every iteration of the algorithm $X_{(n-1) \cdot \bfone} = 0$.
\end{IEEEproof}

In order to reconstruct $W\in\Sigma_q^{[n]^d\setminus\{ (n-1) \cdot \bfone \}}$ from $ X $, the output of Algorithm~\ref{alg:zero-cubes}, we repeatedly inverse the encoding loop. Note that at each iteration that the algorithm encoded a position of an $ L $-cube at Step~\ref{step:encode}, we have $X_{(n-1) \cdot \bfone} = 0$. Thus, we execute the following procedure.
\begin{algorithm}
	\caption{Zero $ L $-Cubes Free Decoding}\label{alg:zero-cubes-dec}
	\begin{algorithmic}[1]
		\While{$X_{(n-1) \cdot \bfone} = 0$}
		\State{Extract $ \bfv \in [n-L]^d $ from $ SD(X_{[n-L,n-1]^d}) $}
		\State{Set  $ A = (\bfv+[L]^d) \cap [n-L,n-1]^d  $}
		\State{Set $X_{[n-L,n-1]^d \setminus A}$ with $ MD(SD(X_{\bfv+[L]^d \setminus A})) $}
		\State{Set $X_{\bfv+[L]^d \setminus A} = \0$}				
		\EndWhile
		\State{Return $ W = X_{[n]^d\setminus\{ (n-1) \cdot \bfone \}} $}
	\end{algorithmic}
\end{algorithm} 

We conclude this section with the following theorem. 
\begin{theorem}
The time complexity of Algorithm~\ref{alg:zero-cubes} and Algorithm~\ref{alg:zero-cubes-dec} is $ \Theta(d  n^d\log (n) ) $.
\end{theorem}
\begin{IEEEproof}
	Both the encoding and decoding algorithms have the same number of iterations, which is $ O(n^d) $. The complexity of the encoding or decoding of $B_{[n]^d}(\bfv)$ for some $ \bfv \in [n]^d $ is $ \Theta(d\log n) $. The actions of reading and writing $ L $-cubes have complexity of $\Theta(L^d)=\Theta(d\log (n))$ as well. Therefore, the time complexity of both algorithms is $ O(d  n^d\log (n) ) $.
\end{IEEEproof}

Lastly, we note that Algorithm~\ref{alg:zero-cubes} works also for the one-dimensional case, which achieves the same value of $L$ as the one achieved by the algorithm presented in~\cite{LevYaa18}. However the complexity of the algorithm in~\cite{LevYaa18} is $\Theta(n)$ while the complexity of Algorithm~\ref{alg:zero-cubes} for the one dimensional case is $\Theta(n\log(n))$.

\begin{example}
	Let $ n = 7, L=3 $, and the input array is
	\[ W = 
	\begin{tikzpicture}[baseline=-\the\dimexpr\fontdimen22\textfont2\relax ]
		\matrix (m)[matrix of math nodes, nodes in empty cells, left delimiter=(,right delimiter=), nodes={text width={width(0111)}, align=center} ]
		{
			0 & 0 & 1 & 0 & 0 & 0 & 1    \\
			0 & 0 & 0 & 0 & 1 & 1 & 1  \\
			0 & 0 & 0 & 0 & 0 & 0 & 0   \\
			0 & 0 & 0 & 0 & 0 & 0 & 0   \\
			0 & 0 & 0 & 0 & 0 & 0 & 1  \\
			0 & 1 & 0 & 0 & 1 & 0 & 0  \\
			0 & 0 & 0 & 1 & 0 & 0 &   \\
		} ;
		
	\end{tikzpicture}.
	\]
The algorithm appends \emph{one} at the missing entry to initialize $ X $. Then, it iterates the coordinates in an increasing order until a zero $ 3 $-square is found. In the following figures, the lookup-square and the found zero square are highlighted.
	\[ X = 
\begin{tikzpicture}[baseline=-\the\dimexpr\fontdimen22\textfont2\relax ]
	\matrix (m)[matrix of math nodes, nodes in empty cells, left delimiter=(,right delimiter=), nodes={text width={width(0111)}, align=center} ]
	{
		0 & 0 & 1 & 0 & 0 & 0 & 1    \\
		0 & 0 & 0 & 0 & 1 & 1 & 1  \\
		0 & 0 & 0 & 0 & 0 & 0 & 0   \\
		0 & 0 & 0 & 0 & 0 & 0 & 0   \\
		0 & 0 & 0 & 0 & 0 & 0 & 1  \\
		0 & 1 & 0 & 0 & 1 & 0 & 0  \\
		0 & 0 & 0 & 1 & 0 & 0 & 1  \\
	} ;
	\draw[color=black, fill=gray, fill opacity=0.4, rounded corners] (m-5-5.north-|m-7-5.west) rectangle (m-7-7.east|-m-7-7.south);
	\draw[color=red, fill=red, fill opacity=0.4, rounded corners] (m-2-1.north-|m-4-1.west) rectangle (m-4-3.east|-m-4-3.south);
	
\end{tikzpicture}
\]
A zero $ 3 $-square found in $ \bfv = (1,0) $. It is replaced with the lookup-square, and the latter is filled with the encoding of $\bfv+\bfe_2 = (1,0) +(0,1) = (1,1)$ using six bits, which is $ B((1,1)) = 000100 $, and appending three more zeros to have a 3-square.
\[ X = 
\begin{tikzpicture}[baseline=-\the\dimexpr\fontdimen22\textfont2\relax ]
	\matrix (m)[matrix of math nodes, nodes in empty cells, left delimiter=(,right delimiter=), nodes={text width={width(0111)}, align=center} ]
	{
		0 & 0 & 1 & 0 & 0 & 0 & 1    \\
		0 & 0 & 1 & 0 & 1 & 1 & 1  \\
		1 & 0 & 0 & 0 & 0 & 0 & 0   \\
		0 & 0 & 1 & 0 & 0 & 0 & 0   \\
		0 & 0 & 0 & 0 & 0 & 0 & 0  \\
		0 & 1 & 0 & 0 & 1 & 0 & 0  \\
		0 & 0 & 0 & 1 & 0 & 0 & 0  \\
	} ;
	\draw[color=black, fill=gray, fill opacity=0.4, rounded corners] (m-5-5.north-|m-7-5.west) rectangle (m-7-7.east|-m-7-7.south);
	\draw[color=red, fill=red, fill opacity=0.4, rounded corners] (m-3-4.north-|m-5-4.west) rectangle (m-5-6.east|-m-5-6.south);
\end{tikzpicture}
\]
Next, a zero $ 3 $-square found in $ \bfv = (2,3) $. It intersects with the lookup-square at positions $ A = \{ (4,4), (4,5) \} $. Hence, the algorithm fills only the non-intersecting part  of $ X_{(2,3) + [3]^2} $ with the non-intersecting portion of the lookup-square, $ \bfy = 0100000 $.
The lookup-square is filled with the encoding of $\bfv+\bfe_2 = (2,3) +(0,1) = (2,4)$, that is, $ B((2,4)) = 010010 $, appended by zeros. 
\[ X = 
\begin{tikzpicture}[baseline=-\the\dimexpr\fontdimen22\textfont2\relax ]
	\matrix (m)[matrix of math nodes, nodes in empty cells, left delimiter=(,right delimiter=), nodes={text width={width(0111)}, align=center} ]
	{
		0 & 0 & 1 & 0 & 0 & 0 & 1    \\
		0 & 0 & 1 & 0 & 1 & 1 & 1  \\
		1 & 0 & 0 & 0 & 1 & 0 & 0   \\
		0 & 0 & 1 & 0 & 0 & 0 & 0   \\
		0 & 0 & 0 & 0 & 0 & 1 & 0  \\
		0 & 1 & 0 & 0 & 0 & 1 & 0  \\
		0 & 0 & 0 & 1 & 0 & 0 & 0  \\
	} ;
	\draw[color=black, fill=gray, fill opacity=0.4, rounded corners] (m-5-5.north-|m-7-5.west) rectangle (m-7-7.east|-m-7-7.south);

\end{tikzpicture}
\]
The algorithm finishes iterating the entries of $ X $ without finding additional zero $ 3 $-squares. The result is indeed a zero $ 3 $-square free array. 
\end{example}
	
	\section{The Cubes Unique Constraint}\label{sec:no-identical-squares}
	In this section, we analyze the size of the set $\cD_{d,q}(n,L)$ and find a condition on $L$ such that its redundancy is at most a single symbol. Furthermore, we provide encoding and decoding schemes for the binary two-dimensional case that use a single redundancy bit.

First, we use a union bound argument to derive a lower bound for $ L $ which assures that the redundancy of $\cD_{d,q}(n,L)$ is at most a single symbol.
\begin{theorem}\label{lem:lcubes-unique-red}
For $L \geq \sqrt[d]{2d \log_q (n) + \log_q(\frac{q}{q-1})} $, it holds that $ |\cD_{d,q}(n,L)| \ge q^{n^d-1} $.
That is, $ \emph{\textmd{red}} (\cD_{d,q}(n,L) )\leq 1$. 
\end{theorem}
\begin{IEEEproof}
If an array $ W \in \Sigma^{[n]^d} $ is not $ L $-cubes unique, then it contains at least two identical $ L $-cubes. The number of possible selections of the identical $ L $-cubes coordinates is bounded from above by $ n^{2d} $. These coordinates can be intersecting or not; in both cases one of the cubes is determined from picking the rest of the $ n^d-L^d $ entries of $ W $. Hence, according to the union bound, the number of arrays over $ \Sigma^{[n]^d} $ that are not $ L $-cubes unique can be bounded from above by
\begin{align*}
n^{2d} q^{n^d-L^d} = q^{n^d} \cdot \frac{n^{2d}}{q^{L^d}} &\leq q^{n^d} \cdot \frac{n^{2d}}{q^{2d \log_q(n)} \cdot q^{ \log_q(\frac{q}{q-1})} } \\&= (q-1)q^{n^d-1},
\end{align*}
where the last inequality follows from the lower bound on $L$. This accordingly implies that $$ |\cD_{d,q}(n,L)| \ge q^{n^d} - (q-1)q^{n^d-1} = q^{n^d-1}. $$ 
\end{IEEEproof}

Next, we present a generic encoding algorithm that uses a single redundancy bit in order to encode binary $ n $-squares that are $ L $-squares unique, for $$ L = 2 \ceilenv{\sqrt{\ceilenv{3 \log (n)} +2}}. $$
Note that this value of $ L $ is far from the value derived in Lemma~\ref{lem:lcubes-unique-red} only by roughly a factor of $ \sqrt{3} $. For simplicity, we sometimes omit ceiling notations in the rest of this section. 

We introduce first a new type of two-dimensional arrays, denoted as \emph{bottom semi squares}, or \emph{semi squares} in short. For a vector $ \bfv \in [n]^2 $, the set $ A =  [n]^2 \setminus (\bfv + [n]^2)$ contains coordinates of a semi square with a corner at $ \bfv $. Hence, we say that $ X \in \Sigma_q^{A} $ is an $ (n,\bfv) $-semi square. 

Let $ X  $ be an $ (n,\bfv) $-semi square for $ \bfv \in [n]^2 $, let $ t$ be an integer, and let $ Y $ be a $ (t,\bfu) $-semi square for $ \bfu \in [t]^2 $. We denote by $ X \circ Y $ the concatenation of $ X $ and $ Y $ which is defined by placing $ Y $ at position $ \bfv $ of $ X $, and restricting the result to the coordinates in $ [n]^2 $.
It follows that $ X \circ Y $ is a $ (n,\bfv+\bfu) $-semi square if and only if for every $ i \in [2] $, $ u_i = 0 $ or $ v_i + t  \ge n$.


\begin{example}\label{ex:semi1}
	\[
	X = 	\begin{pmatrix}
	1 & 1 & 0 & 0 & 1\\
	1 & 0 & 1 & 1 & 1\\
	0 & 0 & 0 & 1 & 0\\
	0 & 0 & \\
	0 & 1 & \\
	\end{pmatrix}
	\in \Sigma^{[5]^2 \setminus ((3,2) + [5]^2)}	
	\]
	is a $ (5,\bfv) $-semi square for $ \bfv = (3,2) $, and
	\[
	Y = 	\begin{pmatrix}
	0 & 0 & 1\\
	1  \\
	0 \end{pmatrix}
	\in \Sigma^{[3]^2 \setminus ((1,1) + [3]^2)}
	\]
	is a $ (3,\bfu) $-semi square for $ \bfu = (1,1) $. Then, the concatenation $ X \circ Y $ is a semi $ (5,\bfv+\bfu) $-square,
	\[
	X \circ Y = 	\begin{pmatrix}
	1 & 1 & 0 & 0 & 1\\
	1 & 0 & 1 & 1 & 1\\
	0 & 0 & 0 & 1 & 0\\
	0 & 0 & 0 & 0 & 1\\
	0 & 1 & 1
	\end{pmatrix}.
	\]
	
\end{example}

\begin{definition}
	Let $ X $ be an $ (n,\bfv) $-semi square for $ \bfv \in [n]^2 $, such that $ \bfv \neq \0 $. We denote by $ CR(X) $ the iterative self-concatenation of $ X $ to an $ n $-square, that is, $$ CR(X) =  X^{\ceilenv{\frac{n}{v_{\min}}}}, $$ where $ v_{\min} $ is the smallest entry of $ \bfv $ that is not $ 0 $.
\end{definition}

It can be shown by induction that after $ m $ concatenations, $ X^m $ is a $ (n,m \cdot \bfv) $-semi square, since $ v_i + n \ge n $ for every $ i \in [2] $. Thus, the self-concatenation of an $ (n,\bfv) $-semi square for every $ \bfv \neq \0 $ is defined properly, and in fact an $ n $-square since for every $ i \in [2] $, $ \ceilenv{\frac{n}{v_{\min}}} \cdot v_i \ge n $. 

\begin{example}\label{ex:semi2}
	Let $ X,Y $ from Example~\ref{ex:semi1}. Then,
	\[
	CR(X) = 	\begin{pmatrix}
	1 & 1 & 0 & 0 & 1\\
	1 & 0 & 1 & 1 & 1\\
	0 & 0 & 0 & 1 & 0\\
	0 & 0 & 1 & 1 & 0\\
	0 & 1 & 1 & 0 & 1\\
	\end{pmatrix}		
	,CR(Y)  = 	\begin{pmatrix}
	0 & 0 & 1 \\
	1 & 0 & 0  \\
	0 & 1 & 0 
	\end{pmatrix}.
	\]	
\end{example}
Additionally, we define the matching \emph{upper semi squares}, described by a coordinate vectors set of $ A =  [n]^2 \setminus (\bfv - [n]^2)$ for $ \bfv \in [n]^2 $. We similarly define concatenation and self-concatenation to an $ n $-square of upper semi squares. 

Algorithm~\ref{alg:no-indentical-squares} receives a two-dimensional array $W \in \Sigma^{[n]^2 \setminus\{(0,0)\}}$, an $ n $-square with a single missing entry, and outputs an $ n $-square $ X \in \cD_{2}(n,L)$.  The algorithm consists of two main procedures, elimination and expansion. First, we initialize $ X $ with $ W $ and set $ 0 $ at the missing entry to mark the start of the elimination. Then, we append to $ X $ a marker $ (L/2) $-square that will mark the transition between the elimination and the expansion parts of the encoder. At the elimination part, we iteratively shorten $ X $ by an $ (L/2) $-square at a time by eliminating one of the two occurrences in $ X $: 1. two identical $ L $-squares, 2. two identical rectangles of size $ [L/2] \times [L] $ (notated for the rest of this section as $ (L/2,L) $-rectangles) where one of them is at the bottom of $ X $. Likewise, we make sure that the marker $ (L/2) $-square appears only once in $ X $. Later, at the expansion part, we enlarge $ X $ to an $ n $-square by iteratively appending $ (L/2) $-squares while making sure that no new identical $ L $-squares are created. 

For convenience, we denote for the rest of this section $$ k = L/2 =\ceilenv{\sqrt{3 \log (n) +2}}.$$ 
We define the \emph{marker $ k $-square} denoted as $ P_M$ as the following square, 
$$ P_M = \begin{pmatrix}
1 & 0 & \cdots & 0  \\
0 & &  &  \\
\vdots &  &  \text{\Huge0}  &  \\
0 &  &  &   \\
\end{pmatrix} 	\in \Sigma^{[k]^2}. $$
Before presenting Algorithm~\ref{alg:no-indentical-squares}, we explain the notion of removal and insertion of squares with granularity. We assume that $n \bmod k = 0 $, and let $ X \in \Sigma^{[n]^2 \setminus A_t} $ where $ A_t $ contains the coordinates of the last $ t $ $ k $-squares in $ [n]^2 $, i.e., $ A_t = ([i_t - k, n-1] \times [j_t, n-1]) \cup ([i_t, n-1]\times [n]) $ where $ (i_t,j_t) = (n-\floorenv{tk / n}, n - (tk \mod n)) $.
We look at $ X $ as a grid of $ k $-squares, and allow only removals and insertions in granularity of $ k $-square units. Removal or insertion actions on a $ k $-square at an aligned position $ (\hat{i}\cdot k,\hat{j} \cdot k) $ are performed by transforming $ X $  to a $ k $-rows array with coordinates $ [k] \times [|X|/k] $, executing the action on the $( i\cdot k + j )$-th $ k $-square like in a one-dimensional array, and transforming back to a grid of $ k $-squares. 

However, during the elimination part of the algorithm we sometimes need to remove a $ k $-square from an unaligned position $ (i,j) $. Such an action is done by finding the closest aligned position $(\hat{i}\cdot k,\hat{j} \cdot k) $, then replacing the data of the non-intersecting parts of $ X_{(\hat{i}\cdot k,\hat{j} \cdot k) + [k]^2} $ and $ X_{(i,j) + [k]^2} $, and finally removing the aligned $ k $-square at position $ (\hat{i}\cdot k,\hat{j} \cdot k) $. This is a technical procedure that can be transparent to the reader of the encoder in Algorithm~\ref{alg:no-indentical-squares}. Nonetheless, for completeness of the encoder, this procedure is explained in Algorithm~\ref{alg:remove-gran}.
\begin{algorithm}
	\caption{Removing a $ k $-Square with Granularity}\label{alg:remove-gran}

	\begin{algorithmic}[1]
		\Statex{Removing a $ k $-square from position $ (i,j) $ of $ X \in \Sigma^{[n]^2 \setminus A_t} $}
		\State{Find maximal $ \hat{i}, \hat{j} $ such that $\hat{i}\cdot k \le i$, $\hat{j}\cdot k \le j$}
		\State{Replace $ X_{(\hat{i}\cdot k,\hat{j} \cdot k) + [k]^2 \setminus ((i,j) + [k]^2)} $ with the data of $ X_{(i,j) + [k]^2 \setminus ((\hat{i}\cdot k,\hat{j} \cdot k) + [k]^2)} $}
		\State{Transform $ \widehat{X} = MD_{[k] \times [|X|/k]}(SD(X)) $}
		\State{Remove $ (\hat{i}\cdot k + \hat{j}) $ $ k $-square from  $ \widehat{X}$}
		\State{Transform $ X = MD_{[n]^2 \setminus A_{t+1}}(SD(\widehat{X})) $}
	\end{algorithmic}
\end{algorithm}

All of the above ensures that appended $ k $-squares, and specifically the marker $ k $-square, are not trimmed or modified accidentally as a result of unrelated removal of insertion actions.

\begin{algorithm*}
	\caption{$ L $-Squares Unique Encoding}\label{alg:no-indentical-squares}
	\algorithmicrequire{ A two-dimensional array $W \in \Sigma^{[n]^2 \setminus\{(0,0)\}}$ }
	\\ \algorithmicensure{ An $ n $-square $X \in \cD_{2}(n,L)$}
	\begin{algorithmic}[1]
		\State{Set a square $ X \in \Sigma^{[n]^2} $ with $ X_{0,0} = 0, X_{[n]^2 \setminus\{(0,0)\}} = W $}\label{step:nis-init1}
		\State{Denote $ (i_m, j_m) = (n,0)$, append $ X_{(i_m,j_m) + [k]^2} = P_M$}\label{step:nis-init2}
		\Statex{First part - Elimination}
		\While{at least one of the occurences in cases 1,2,3 exists}
		\Case{ \textbf{1}: A $ k $-square equals to $ P_M $ exists at $ (i_1,j_1) < (i_m,j_m) $ }
		\State{Remove square $ X_{(i_1,j_1)+[k]^2 } $  }
		\State{Set $ \bfv =  101 \circ B_{[n]^2}(i_1,j_1) \circ 1^{k^2 - 2\log (n) - 3} $, insert $ MD_{[k]^2}(\bfv) $  at $ X_{0,0} $}\label{step:nis-enc3}	
		\EndCase	
		\Case{ \textbf{2}: Identical $L$-squares exist at $ (i_1,j_1) < (i_2,j_2) $}	
		\State{Remove square $ X_{(i_1,j_1)+[L]^2 } $ }
		\State{Set $ \bfv = 100 \circ B_{[n]^2}(i_1,j_1) \circ B_{[n]^2}(i_2,j_2) \circ 1^{3k^2 - 4\log (n) - 3} , $ insert $ MD_{[k]\times[3k]}(\bfv) $  at $ X_{0,0} $}\label{step:nis-enc1}
		\EndCase	
		\Case{ \textbf{3}: Identical $ (k,L) $-rectangles exist at $ (i_1,j_1) < (i_2,j_2) $, where $$ (i_2,j_2) \in I = (\{i_m -k\} \times [j_m - k ,n-1]) \cup (\{ i_m \} \times [j_m - k-1]) , $$\vspace{-3ex}}	\label{step:nis-case2}
		\State{Remove rectangle $ X_{(i_1,j_1)+[k]\times[L] } $  }		
		\State{Set $ \bfv =  11 \circ  B_{[n]^2}(i_1,j_1) \circ B_I(i_2,j_2) $, insert $ MD_{[k]^2}(\bfv) $ at $ X_{0,0} $}\label{step:nis-enc2}
		\EndCase				
		\State{If cases 2 or 3 were executed, decrement $ (i_m, j_m) $ by a  $ k $-square}
		\EndWhile \label{step:nis-after-elim}
		\State{If $ |X| \ge n^2 $, return $ X_{[n]^2} $}\label{step:ret1}
		\Statex{ Second part - Expansion}
		\While{$ |X| < n^2 $}
		\State{Set indexes $ (i_e,j_e) $ to point to the next missing $ k $-square in $ X $}
		\State{Let $ I_e = ( (i_e,j_e) - [k]^2 ) \cap [n]^2 $. Set \[
			\cS = \{ X_{(i,j) + [k]^2} \mid (i,j) \notin I_e \} \cup \{CR(X_{(i,j) + [k]^2 }) \mid (i,j) \in I_e \}
			\]}
		\State{Pick $ Y \in \Sigma^{[k^2]}/\cS $ and set $ X_{(i_e,j_e) + [k]^2} = Y $}\label{step:nis-pick}
		
		\EndWhile
		\State{Return $ X $}\label{step:ret2}
	\end{algorithmic}
\end{algorithm*}
We prove the correctness of the algorithm in the next few claims.

	\begin{claim}\label{clm:nis-elim}
	Algorithm~\ref{alg:no-indentical-squares} reaches Step~\ref{step:nis-after-elim}, i.e., the elimination part terminates. Additionally, all assignments executed throughout the elimination are correctly defined. 
\end{claim}	
\begin{IEEEproof}
	First, notice that all removal and insertion actions are done with granularity of $ k $-square units, as allowed. 
	We prove that the elimination loop terminates by showing that at each iteration of the elimination loop, the length of $ X $ decreases or the  Hamming weight of $ X $ increases. We analyze each case of removal and insertion independently. 
	\\ \textbf{Case 1:} We remove a square of size $ k^2 $ with Hamming weight equals to $ w_H(P_M) = 1 $ and insert a square of size $ k^2 $ with Hamming weight of at least $ \log (n) $.	
	\\ \textbf{Case 2:} We remove a square of size $ L^2 = 4k^2 $ and insert a smaller rectangle of size $ 3k^2 $.
	\\ \textbf{Case 3:} We remove a square of size $ L\cdot k = 2k^2 $ and insert a smaller rectangle of size $ 3 \log (n) + 2 = k^2 $. This follows from the fact that the number of possible indices for $ (i_2,j_2) $ satisfies $ |I| \le n $. 
\end{IEEEproof}

	\begin{claim}\label{clm:elim-stuff}
	At Step~\ref{step:nis-after-elim} of Algorithm~\ref{alg:no-indentical-squares}, the two-dimensional array $ X $ satisfies the following properties:
	\begin{enumerate}[(1)]
		\item  $ X $ is $ L $-square unique,
		\item  $ X $ contains no identical $ (k,L) $-rectangles where one of them is at position that belongs to
		$$I = (\{i_m -k\} \times [j_m - k ,n-1]) \cup (\{ i_m \} \times [j_m - k-1]) , $$
		\item  $ X $ ends with $ X_{(i_m, j_m)+[k]^2} = P_M $, 
		\item  $ X $ contains no other $ k $-square equals to $ P_M $.
	\end{enumerate}
\end{claim}

\begin{IEEEproof}
	First, we prove (3) by showing that throughout the elimination loop, $ X_{(i_m, j_m)+[k]^2} = P_M $. This holds at Step~\ref{step:nis-init2} before the elimination loop, and during the elimination loop the indices $ (i_m, j_m) $ are decremented by a $ k $-square if and only if $ X $ was shortened by $ k^2 $. Thus, this condition can be violated only if some part of $ X_{(i_m, j_m)+[k]^2} $ was removed as part of an elimination procedure in cases 2 or 3. 	Assume in the contrary that case 2 occurred and there were two identical squares at positions $ (i_1, j_1) < (i_2,j_2) $ such that $ X_{(i_1,j_1) + [L]^2} $ intersects with $ X_{(i_m,j_m) + [k]^2} $. Thus, $ X_{(i_1,j_1) + [L]^2} $ contains the 1-bit at the top-left corner of $ P_M $ at some position $ (i_r,j_r) $. However, it follows that $ X_{(i_2+i_r,j_2+j_r)} = 1 $ which is a contradiction since $ P_M $ contains a single 1-bit and $ (i_1, j_1)\neq (i_2,j_2) $. It can be shown similarly that a part of $ X_{(i_m, j_m)+[k]^2} $ can not be removed in case 3. Statements (1), (2), (4) follows from the fact that the elimination loop terminates as proved in Claim~\ref{clm:nis-elim}, using case 2, 3, 1, respectively.
\end{IEEEproof}

\begin{claim}
	For every iteration of the expansion loop, the set $\Sigma^{[k]^2} \setminus \cS $ is not empty. 
\end{claim}

\begin{IEEEproof}
	The size of the set $ \cS $ satisfies
	\[
	|\cS| \le |X| \le n^2,
	\]
	while the alphabet $\Sigma^{[k]^2}$ satisfies
	\[
	|\Sigma^{[k]^2} | \geq 2^{3 \log (n) + 2} > n^2.
	\]
\end{IEEEproof}

\begin{figure}
	\includegraphics[width=\columnwidth] {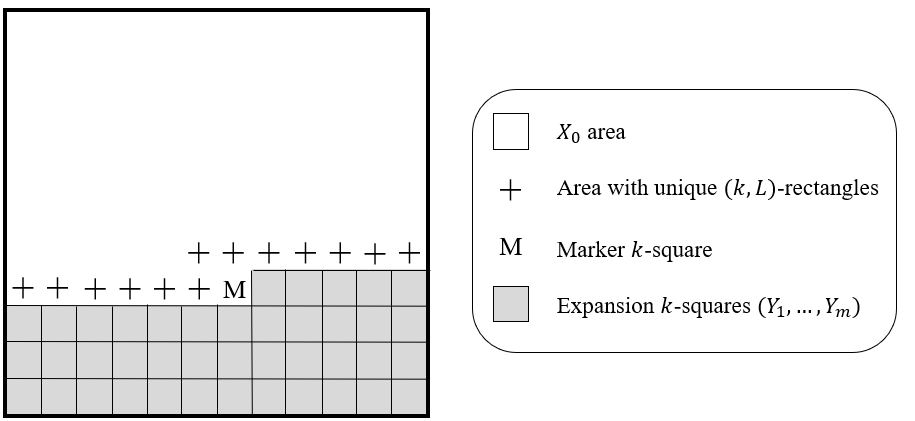}
	\centering
	\caption{Structure of $ X $ at the end of the expansion part}
	\label{fig:nis1}
\end{figure}

Let $ m $ denote the number of iterations of the expansion loop of Algorithm~\ref{alg:no-indentical-squares} that were executed. For every $ \ell = 1, \dots, m $, let $ X_\ell $ denote the value of $ X $ at the end of the $ \ell $-th iteration, and let $ Y_\ell $ denote the expansion $ k $-square that the algorithm picked at Step~\ref{step:nis-pick}. We notate by $ X_0 $ the value of $ X $ before the first iteration of the expansion loop, i.e., at Step~\ref{step:nis-after-elim} right after the elimination part.  Figure~\ref{fig:nis1} presents an example of the structure of $ X $ at the end of the expansion part.

\begin{claim}\label{clm:nis-appear-once}
	For every iteration $ \ell = 1, \dots,m $, the array $ X_\ell = X_{\ell-1} \circ Y_\ell $ contains the square $ Y_\ell $ only once, at its end. 
\end{claim}

\begin{IEEEproof}
	Let $ i_e, j_e $ denote the position of $ Y_\ell $. 
	According to the construction of $ \cS $, the square $ Y_\ell $ can not appear as a sub-square of $ X_{\ell-1} $. Thus, it might appear at some position  $ (i, j) $ of $ X $, where $ (i,j) \in (i_e,j_e) - [k]^2$.
	Assume in the contrary that such a case occurs. We have 
	$$ (X_\ell)_{(i,j) + [k]^2} = (X_{\ell-1} \circ Y_\ell)_{(i,j) + [k]^2} =  Y_\ell. $$ However, it is implied that $$ Y_\ell = CR((X_{\ell-1})_{(i,j) + [k]^2 }),$$ which is a contradiction to the construction of $ \cS $. 
\end{IEEEproof}

	\begin{claim}\label{clm:nis-expand}
	At Step~\ref{step:ret2} of Algorithm~\ref{alg:no-indentical-squares}, $ X $ is $ L $-square unique. 
\end{claim}

\begin{figure}
\includegraphics[width=\columnwidth] {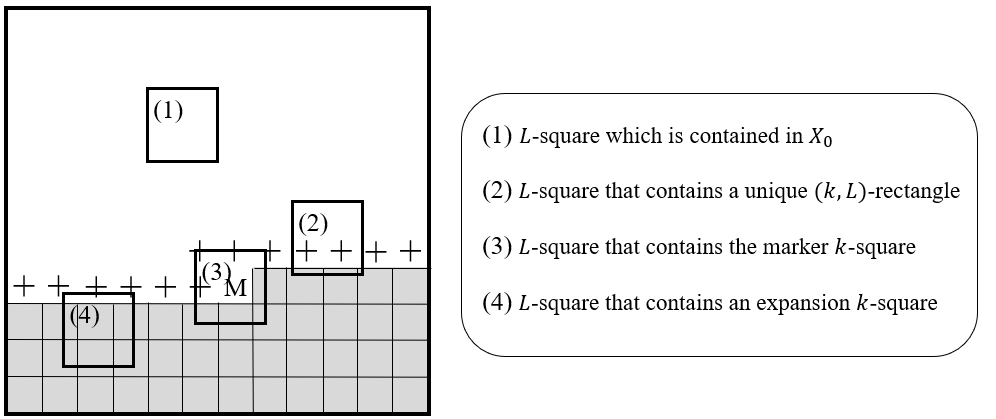}
\centering
\caption{different cases of $ L $-squares for the proof of Claim~\ref{clm:nis-expand}, based on the structure presented in Figure~\ref{fig:nis1}}
\label{fig:nis2}
\end{figure}

\begin{IEEEproof}
	Assume in the contrary that $ X $ contains two identical $ L $-squares at positions $ (i_1,j_1) < (i_2, j_2) $. We prove the claim by examining all different cases for $ (i_2, j_2) $ and reaching a contradiction at each case. These cases are also presented graphically at Figure~\ref{fig:nis2}.
	\begin{enumerate}[(1)]
		\item	If $ X_{(i_2,j_2)+[L]^2} $ is contained in $ X_0 $, we have a contradiction since $ X_0 $ is $ L $-square unique from Claim~\ref{clm:elim-stuff} Statement~(1). 
		\item	If $ X_{(i_2,j_2)+[L]^2} $ contains an $ (k,L) $-rectangle which starts at position which belongs to  $$I = (\{i_m -k\} \times [j_m - k ,n-1]) \cup (\{ i_m \} \times [j_m - k-1]) , $$ 
		it follows that $ X_{(i_1,j_1)+[L]^2}$ contains an identical $ (k,L) $-rectangle which is a contradiction to Claim~\ref{clm:elim-stuff} Statement~(2).
		\item	If $ X_{(i_2,j_2)+[L]^2} $
		contains at some position $ (i_r,j_r) $ the marker $ k $-square $ X_{(i_m,j_m) + [k]^2} $, it follows that $ X_{(i_2+i_r,j_2+j_r)+[k]^2} = P_M$ from  Claim~\ref{clm:elim-stuff} Statement~(3). However, therefore $ X_{(i_1+i_r,j_1+j_r)+[k]^2} = P_M$ as well which is a contradiction to Claim~\ref{clm:elim-stuff} Statement~(4).
		\item	Otherwise, $ X_{(i_2,j_2)+[L]^2}$  contains an expansion $ k $-square at some position $ (i_r,j_r) $. That is, $ X_{(i_2+i_r,j_2+j_r)+[L]^2} = Y_\ell $ where $ Y_\ell $ is a $ k $-square that was appended to $ X_{\ell-1} $ at the $ \ell$-th iteration of the expansion loop. It follows that $ X_{(i_1+i_r,j_1+j_r)+[k]^2} = Y_\ell $ as well. Thus, $ Y_\ell $ appears twice in $ X_\ell $ which is a contradiction to Claim~\ref{clm:nis-appear-once}.
	\end{enumerate}

\end{IEEEproof}		

\begin{lemma}
	Algorithm~\ref{alg:no-indentical-squares} successfully returns an $ n $-square that satisfies the $ L $-squares unique constraint. 
\end{lemma}
\begin{IEEEproof}
	From Claim~\ref{clm:nis-elim}, the elimination loop terminates and the algorithm reaches Step~\ref{step:nis-after-elim}. If the condition in Step~\ref{step:ret1} is satisfied, then $ X_{[n]^2} $ is an $ n $-square and is also $ L $-square unique from Claim~\ref{clm:elim-stuff} Statement~(1). Otherwise, the algorithm reaches Step~\ref{step:ret2} with $ X \in \Sigma^{[n]^2} $ which is $ L $-square unique as well from Claim~\ref{clm:nis-expand}.
\end{IEEEproof}	

The decoding scheme receives $ X $ which is an output of Algorithm~\ref{alg:no-indentical-squares} and returns $ W \in \Sigma^{[n]^2 \setminus\{(0,0)\}} $. First, we identify the marker square position by looking at the first occurrence of $ P_M $. Using Claim~\ref{clm:nis-elim} we can remove the part of $ X $ after the marker square since it was appended during the expansion procedure. Next, we iteratively inverse the elimination procedure. We identify using the first three entries of $ X $ the last elimination case at which data was encoded. If data was encoded at case~2, we decode $ \bfv = SD(X_{[k]\times[3k]}) $, extract the positions $ (i_1,j_1), (i_2,j_2) $ and insert $ X_{(i_2,j_2) + [L]^2} $ at position $ (i_1,j_1) $ if the $ L $-squares do not intersect. Otherwise, we insert instead the self-concatenation $ CR(X_{((i_2,j_2) + [L]^2) \setminus ((i_1,j_1) + [L]^2) }) $. Similarly, in case~3 we recover $ X_{(i_1,j_1) + [k]\times[L]} $ from $ \bfv = SD(X_{[k]^2}) $. If the data was encoded at case~1, we decode $ \bfv =  SD(X_{[k]^2}) $, extract  $ (i_1,j_1)$ and insert $ X_{(i_1,j_1) + [k]^2} = P_M $. This process is repeated until $ X_{(0,0)} = 0 $, then we return $X_{[n]^2 \setminus\{(0,0)\}} $ as $ W $.

\begin{remark}
Algorithm~\ref{alg:no-indentical-squares} requires that $ n \bmod k = 0 $. However, the algorithm can be altered to support cases where this is not possible, e.g. $ n $ is a prime number. In this case, we pad the input array in the right and the bottom with \emph{ones} in order to receive an $ n' $-square, where $ n' \ge n$ is the closest multiple of $ k $ to $n$. Then, we invoke Algorithm~\ref{alg:no-indentical-squares} with minor modifications that are described shortly to receive $ X' \in \cD_2(n',L) $, and return $ X = X'_{[n]^2} $. Some information that is valuable for the decoder can be lost when restricting the result to an $ n $-square. In order for the decoder to uniquely identify the marker-square $ P_M $, we pick at Step~\ref{step:nis-pick} only squares with $ Y_{0,0} = 0 $. Additionally, we make sure that the  area padded with \emph{ones} that is contained in $ X_0 $ (the array $X$ before the expansion part)  remains unchanged throughout the elimination. This can be done by adding two cases that are similar to cases 2 and 3 that are specific to when the identical $ L $-square or $ (k,L) $-rectangle found intersects with the padded area. These new cases will encode the special occurrence with a small number of bits and only the non-intersecting part of the sub-array will be removed. Both these modifications will not change the redundancy of the algorithm.
\end{remark}

	\section{Extensions to Multidimensional Boxes of any Volume}\label{sec:ext-v}
	In this section, we introduce a generalization of the zero $ L $-cubes free and the $ L $-cubes unique constraints to multidimensional arrays where the shape of the sub-array is not necessarily a cube, but a box, and only the volume is given. A $ d $-dimensional box is a shape that generalizes the shape of a rectangle to any dimension $ d $, and is given by a set of coordinates $ A = [x_0] \times \cdots \times [x_{d-1}] $ where the sides $ x_0, \dots, x_{d-1} $ are positive integers that belong to $ [1,n] $. The volume of such a box is given by $ |A| = x_0 \cdots x_{d-1} $.

\begin{definition}\label{def:zero-box}
Let $ W \in \Sigma^{[n]^d} $ be a $ d $-dimensional array. For a positive integer $V$, we say that $W $ contains a \textbf{zero $ V $-box} (or \textbf{zero $ V $-rectangle} for $ d=2 $) with a coordinates set $ A = [x_0] \times \cdots \times [x_{d-1}]  $ at position $ \bfv $ such that $ \bfv + A \subseteq [n]^d $, if $ W_{\bfv + A}~=~\0$ and $ |A|=V $. An array $ W $ satisfies the \textbf{zero $ V $-boxes free constraint} if it does not contain a zero $ V' $-box, for any positive integer $ V' \ge V $.
\end{definition}

\begin{definition}\label{def:unique-box}
Let $ W \in \Sigma^{[n]^d} $ be a $ d $-dimensional array. For a positive integer $V$, we say that $W $ contains  \textbf{identical $ V $-boxes} (or \textbf{identical $ V $-rectangles} for $ d=2 $) with a coordinates set $ A = [x_0] \times \cdots \times [x_{d-1}]  $ at positions $ \bfu \neq \bfv $ such that $ \bfu~+~A,\bfv~+~A \subseteq [n]^d $, if $ W_{\bfu + A}  = W_{\bfv + A}$ and $ |A|=V $. An array $ W $ satisfies the \textbf{$ V $-boxes unique constraint} if it does not contain two identical $ V' $-boxes, for any positive integer $ V' \ge V$.
\end{definition}
In the rest of this paper, we sometimes refer to an array that satisfies the constraint in Definition~\ref{def:zero-box} as a \emph{zero $ V $-boxes free} array, and to an array that satisfies the constraint in Definition~\ref{def:unique-box} as a \emph{$ V $-boxes unique} array.
We denote by $ \cC\cA_{d,q}(n,V) $ the set of all arrays over $ \Sigma^{[n]^d} $ that satisfy the zero $ V $-boxes free constraint and by $ \cD\cA_{d,q}(n,V) $ the set of all arrays over $ \Sigma^{[n]^d} $ that satisfy the  $ V $-boxes unique constraint.

\subsection{Enumeration of Minimal Boxes}

Before analyzing the constraints, we first need to estimate the number of \emph{minimal boxes} for a given volume. For an integer $ V $, Let $ F_d(V) $ denote the set of minimal boxes  $ A = [x_0] \times \cdots \times [x_{d-1}]  $ such that $ |A| \ge V $ and for every other $ A' \in F_d(V), A \not \subset A' $. Additionally, let $ f_d(V) = |F_d(V)| $.

First, we examine $ f_d(V) $ for small values of $ d $. It is clear that $ f_1(V) = 1$. For $ d = 2$, there are $ \lfloor{\sqrt{V}}\rfloor $ possibilities for the smaller side of the rectangle, which yields 
$$ f_2(V) = \begin{cases}
	 2\lfloor\sqrt{V}\rfloor & \sqrt{V} \not \in \mathbb{N}\\
	 2\sqrt{V} -1 & \sqrt{V} \in \mathbb{N}\\
\end{cases},  $$
since when $ \sqrt{V} $ is an integer, the cube $ [\sqrt{V}]^2 $ is counted twice. When it is not known if $ \sqrt{V} $ is an integer, we can write $ f_2(V) \le 2\sqrt{V} $.

After acquiring some intuition regarding the value of $ f_d(V) $,
we have the following claim in the general case.
\begin{claim}\label{clm:num-boxes-V}
For every $ d \ge 2 $ and for every positive integer $ V $, $$ f_d(V) \le  \alpha^{d-2} d!(d-1)! V^{\frac{d-1}{d}},  $$
where $ \alpha $ is a constant that satisfies $ 1 \le \alpha \le \sqrt{2}  $ and approaches~$ 1 $ as  $ V \rightarrow \infty $.
\end{claim}
\begin{IEEEproof}
We can write $ f_d(V) $ as a recursive inequality,
$$  f_d(V) \le d\sum_{x=1}^{\lfloor \sqrt[d]{V} \rfloor} f_{d-1}\big(\ceilenv{\frac{V}{x}}\big),  $$
which follows from having $ d $ options for the shortest side, that is at most $ \lfloor \sqrt[d]{V} \rfloor $, and letting the remained $ d-1 $ coordinates determine the volume of the box.
Let $ \alpha > 1 $ be the minimal constant that satisfies $ \lceil \frac{V}{x} \rceil \le \frac{\alpha V}{x} $ for every $ x \le \lfloor \sqrt{V} \rfloor $. It follows that $ \alpha \le \frac{\lceil \sqrt{V} \rceil}{\sqrt{V}} \le 1 + \frac{1}{\sqrt{V}}$. Thus, $ \alpha \le \sqrt{2} $ from plugging $ V=2 $, and $ \limup{V} \alpha \le \limup{V} 1 + \frac{1}{\sqrt{V}} = 1 $.
We prove the claim statement inequality using an induction. 
From previous calculations, $ f_2(V) \le 2 \sqrt{V} $ which verifies the claim's inequality for $d=2$. 

Assume next that the inequality holds for $ d-1 $. Then, we have that 
\begin{equation*}
\begin{aligned}
f_d(V) &\le d\sum_{x=1}^{\lfloor \sqrt[d]{V} \rfloor} f_{d-1}\big(\ceilenv{\frac{V}{x}}\big) 
\\&\le  d\sum_{x=1}^{\lfloor \sqrt[d]{V} \rfloor} f_{d-1}\big(\frac{\alpha V}{x}\big)
\\&\le d\sum_{x=1}^{\lfloor \sqrt[d]{V} \rfloor} \alpha^{d-3}(d-1)!(d-2)! \big(\frac{\alpha V}{x}\big)^{\frac{d-2}{d-1}}
\\&\overset{(a)}{\le} \alpha^{d-2} d!(d-2)!  V^{\frac{d-2}{d-1}} \int_{x=0}^{\sqrt[d]{V}} \frac{dx}{x^{\frac{d-2}{d-1}}}
\\&=\alpha^{d-2}  d!(d-2)! V^{\frac{d-2}{d-1}} \left[ (d-1)x^{\frac{1}{d-1}} \right]_{0}^{\sqrt[d]{V}}
\\&\le  \alpha^{d-2}  d!(d-1)!V^{\frac{d-2}{d-1}} V^{\frac{1}{(d-1)d}} 
\\&=  \alpha^{d-2} d!(d-1)! V^{\frac{d-1}{d}} 
\end{aligned}
\end{equation*}
where (a) from the inequality $ \sum_{i=L}^{U} g(i) \le \int_{L-1}^{U} g(x)dx$ for a nonnegative decreasing function $ g(x) $.
\end{IEEEproof}

\begin{claim}\label{clm:num-boxes-V-lower}
For every $ d \ge 2 $ and for every positive integer $V$, $$ f_d(V) \ge d (\lfloor \sqrt[d]{V} \rfloor)^{d-1} -d + 1.$$
\end{claim}
\begin{IEEEproof}
	Let $ S $ denote the set of $ d $-dimensional boxes that are generated by letting the first $ d-1 $ sides have any value of $ [1,\lfloor \sqrt[d]{V} \rfloor] $, and letting the last side fill the remaining volume of the box to $ V $. That is,
	\begin{align*}
		S  =\left\{ [x_0] \times \cdots \times [x_{d-1}]\middle\vert 
		\begin{array}{l}
			x_0, \dots, x_{d-2} \in  [1,\lfloor \sqrt[d]{V} \rfloor], \\
			x_{d-1} = \ceilenv{\frac{V}{x_0 \cdots x_{d-2}}} .
		\end{array}\hspace{-0.5ex}\right\}
	\end{align*}
	Let $ A_0 = [x_0] \times \cdots \times [x_{d-1}] \in S $ be a box. Clearly, $ |A_0| \ge V $. Hence, in order to prove that $ S \subseteq F_d(V) $, it is left to show that $ S $ is minimal. Assume that there exists $ A_1 = [y_0] \times \cdots \times [y_{d-1}] \in S $ and w.l.o.g there exists $i \in [d-1] $ such that $ y_i > x_i $ and for every other $ j \in [d-1]\setminus\{i\} $, $ y_j \ge x_j $. It follows that 
	\begin{align*} \frac{V}{x_0 \cdots x_{d-2}} &- \frac{V}{y_0 \cdots y_{d-2}} \\&\ge \frac{V}{y_0 \cdots y_{i-1} y_{i+1} \cdots  y_{d-2} x_i} - \frac{V}{y_0 \cdots y_{d-2}} 
		\\&= \frac{V(y_i - x_i)}{y_0 \cdots y_{d-2}x_i} 
		\\&\ge \frac{V(y_i - x_i)}{(\sqrt[d]{V})^d} 
		\\&= y_i - x_i 
		\\&\ge 1.
		\end{align*}
	Therefore $ x_{d-1} > y_{d-1} $ and hence $ A_0 \not \subset A_1 $. Thus, $ S $ is minimal. 
	
	From its definition, we have that $ |S| = (\lfloor \sqrt[d]{V} \rfloor)^{d-1} $. If $ {\sqrt[d]{V} \not \in \mathbb{N}} $, we can shift $ d-1 $ times the sides of each $ A \in S $  in order to generate additional unique boxes that belong to $ F_d(V) $. This holds since the remaining side satisfies $ x_{d-1} \ge \lceil\sqrt[d]{V}\rceil $ where the other sides are at most $\lfloor \sqrt[d]{V} \rfloor$. In the case where $ \sqrt[d]{V} \in \mathbb{N} $ we can shift each box of $ S $ but the set $ [\sqrt[d]{V}]^d $. We can conclude that $$ f_d(V) \ge d (\lfloor \sqrt[d]{V} \rfloor)^{d-1} -d + 1. $$
%
%
\end{IEEEproof}
The next corollary follows immediately from Claims~\ref{clm:num-boxes-V} and~\ref{clm:num-boxes-V-lower}.
\begin{corollary}\label{cor:fdv}
For every fixed positive $ d \in \mathbb{N} $ and for every positive $ V $, $$ f_d(V) = \Theta(V^{\frac{d-1}{d}}). $$
\end{corollary}
Even though for the results in the paper only an upper bound on the value of $f_d(V)$ would be sufficient, we still found it important to present Corollary~\ref{cor:fdv} for a more comprehensive analysis of the value of $f_d(V)$. In particular, for every $ d \in \mathbb{N} $, we can write  $f_d(V) \leq C_d V^{\frac{d-1}{d}}$ where $ C_d $ denotes a positive constant that fulfills Corollary~\ref{cor:fdv}. Note that regarding our constraints, the number of minimal boxes of volume $ V $ that are contained in $ W \in \Sigma_q^{[n]^d} $ and start at position $ \bfv $ depends on $n$ and $ \bfv $ in addition to the volume $ V $. However, it can be bounded from above by $f_d(V)$.

\subsection{The Zero Boxes Free Constraint}
First, we prove the following lemma regarding the cardinality of  the set $ \cC\cA_{d,q}(n,V) $.
\begin{lemma}\label{lem:vboxes-free-red2}
	For $ V = d \log_q(n) + \frac{d-1}{d}\log_q(\log_q(n))+ \cO(1) $, and for $ n $ large enough it holds that $ |\cC\cA_{d,q}(n,V)| \ge q^{n^d-1} $. That is, $ \ered{\cC\cA_{d,q}(n,V)} \le 1$ . 
\end{lemma} 
\begin{IEEEproof}
	Let $ V = d \log_q(n) + \frac{d-1}{d}\log_q(\log_q(n)) + C + \log_q(\frac{q}{q-1}) $ for some positive constant $ C $ that will be determined later. 
	If an array $ W \in \Sigma^{[n]^d} $ is not zero $ V $-boxes free, then it contains at least a single zero box with coordinates set $ A \in F_d(V) $, such that $ |A| \ge V $. From Corollary~\ref{cor:fdv} there are at most $ C_d V^{\frac{d-1}{d}} $ possible selections of such a coordinates set. Hence, according to the union bound, the number of arrays that are not zero $ V $-boxes free can be bounded from above by
	\begin{equation}\label{eq:vboxes-free-red2}
	\begin{aligned}	
	n^{d} C_d V^{\frac{d-1}{d}} &\cdot q^{n^d-V} = q^{n^d} \cdot \frac{n^{d} C_d V^{\frac{d-1}{d}}}{q^{V}}
	\\&= (q-1)q^{n^d-1} \cdot \frac{C_d V^{\frac{d-1}{d}}}{q^{\frac{d-1}{d}\log_q(\log_q(n)) + C}}
	\\&\overset{(a)}{\le}  (q-1)q^{n^d-1} \cdot \frac{C_d( (d+1)\log_q(n))^{\frac{d-1}{d}}}{\log_q(n)^{\frac{d-1}{d}}q^C}
	\\&\overset{(b)}{\le} (q-1)q^{n^d-1} .
	\end{aligned}
	\end{equation}
	Inequality (a) follows from $ V \le (d+1)\log_q(n) $ for $n$ large enough and (b) holds  by choosing $ C \ge \log_q(C_d (d+1)^{\frac{d-1}{d}}) $.
	This accordingly implies that $ |\cC\cA_{d,q}(n,V)| \ge q^{n^d-1} $. 
\end{IEEEproof}

When comparing the result of Lemma~\ref{lem:vboxes-free-red2} with the lower bound derived in Theorem~\ref{th:red-cdqnL} for arrays that are zero $ L $-cubes free, it follows that for the same volume $ V = L^d $, the minimal volume required for a redundancy of one symbol in the latter case is smaller by $ \Delta = \frac{d-1}{d}\log_q(\log_q(n)) + \cO(1) $.

Next, we present an encoding algorithm that uses a single redundancy symbol to encode $ V $-boxes free cubes over $ \Sigma_q^{[n]^d}$, for
$$ V =\ceilenv{ d \log_q(n)} + \ceilenv{\frac{d-1}{d}\log_q(\log_q(n))}+ C + 1, $$
where $ C = \ceilenv{\log_q(C_d) + \frac{d-1}{d}\log_q(d+1)} $, i.e., the ceiling of the constant from the proof of Lemma~\ref{lem:vboxes-free-red2}. Note that this value of $ V $ adds at most four redundancy symbols to the lower bound derived in the proof of Lemma~\ref{lem:vboxes-free-red2}. For simplicity, we  omit the ceiling notation in the rest of this section.

Algorithm~\ref{alg:vboxes-free} receives a $d$-dimensional array $W\in\Sigma^{[n]^d\setminus\{ \0 \}}$ with a single symbol missing, and outputs a cube $X \in \cC\cA_{d,q}(n,V)$. First, we initialize $ X $ with $ W $ and set 0 at the missing entry to mark the start of the algorithm. Then, we iteratively look for zero $ V $-boxes in $ X $. When such a box is found, we remove it from $ X $, and insert at the beginning  of $ X $ an encoding of the position and the shape of the box, along with additional $ 1 $-bits. Thus, we ensure that the Hamming weight of the square increases and the algorithm eventually terminates.

The insertions and deletions in this algorithm are preformed with granularity of $ 1 $, i.e., as a one-dimensional sequence. In particular, at Step~\ref{step:vboxes-remove} we remove from $ X $ a box with coordinates $ A $ at position $ \bfu $ by performing $ \bfx = SD(X) $ and removing from $ \bfx $ the entry at position $ \sum_{i=0}^{d-1}n^{i} u'_{d-i}  $ for every $ \bfu' = (u'_1, \dots, u'_d) \in \bfu + A $. Then at Step~\ref{step:vboxes-insert} we insert a length-$ |A| $ vector at the beginning of $ \bfx $ and retransform it to a cube by $ X = MD_{[n]^d}(\bfx) $. 

\begin{algorithm}
	\caption{Zero $ V $-Boxes Free Encoding}\label{alg:vboxes-free}
	\algorithmicrequire{ A $ d $-dimensional array $W \in \Sigma^{[n]^d \setminus\{\0\}}$ }
	\\ \algorithmicensure{ A $ d $-dimensional array $X \in \cC\cA_{d,q}(n,V)$}
	\begin{algorithmic}[1]
		\State{Set an array $ X \in \Sigma^{[n]^d} $ with $ X_{\0} = 0, X_{[n]^d \setminus\{\0\}} = W $}\label{step:vboxes-init1}
		
		\While{there exists a zero box $ X_{\bfu+A} = \0 $ where $ A \in F_d(V) $}\label{step:vboxes-while}
		\State{Remove box $X_{\bfu+A} $}\label{step:vboxes-remove}
		\State{Set $ \bfv = 1 \circ B_{[n]^d}(\bfu)  \circ b_{F_d(V)}(A) $}
		\State{Insert $ \bfv \circ 1^{|A|-|\bfv|} $ at the start of $ X $} \label{step:vboxes-insert}
		\EndWhile	
		\State{Return $ X $}\label{step:vboxes-ret}
	\end{algorithmic}
\end{algorithm}

\begin{lemma}
	Algorithm~\ref{alg:vboxes-free} successfully outputs a $ d $-dimensional array that satisfies the zero $ V $-boxes free constraint.
\end{lemma}

\begin{IEEEproof}
	First, the assignment in Step~\ref{step:vboxes-insert} is correctly defined since on one hand, 
	\begin{align*}
|A| \ge V = d\log_q(n) + \frac{d-1}{d}\log_q(\log_q(n))+ C + 1,
	\end{align*}
and on the other hand,
	\begin{align*}
	|\bfv| &\le d \log_q(n) + \log_q(f_d(V)) + 1 
	\\&\le d \log_q(n) + \log_q(C_d V^{\frac{d-1}{d}}) + 1
	\\& = d \log_q(n) + \log_q(C_d) + \frac{d-1}{d}\log_q(V) +1
	\\& \le d \log_q(n) + \log_q(C_d) + \frac{d-1}{d}\log_q((d+1)\log_q(n)) +1
	\\& \le d \log_q(n) +  \frac{d-1}{d}\log_q(\log_q(n)) + C +1
	\\& = V.
\end{align*}
		
	Thus, it follows that throughout the while loop of the algorithm the size of $ X $ remains exactly $ n^d $. Since we remove a box of zeros at Step~\ref{step:vboxes-remove} and insert data with Hamming weight of at least $ 1 $ at Step~\ref{step:vboxes-insert}, the Hamming weight of $ X $ increases at every iteration. Therefore, the while loop eventually stops and the algorithm reaches Step~\ref{step:vboxes-ret}.
	
	Next, assume in the contrary that $ X $, which is returned in Step~\ref{step:vboxes-ret}, is not zero $ V $-boxes free. Thus, $ X $ contains a zero box at position $ \bfu $ and a coordinates set $ A \in F_d(V) $ which contradicts the condition of the loop in Step~\ref{step:vboxes-while}.
\end{IEEEproof}

The decoder reconstructs $W\in\Sigma^{[n]^d\setminus\{ \0 \}}$ from $ X $, an output of Algorithm~\ref{alg:zero-cubes}, by inverting the encoding loop. Note that at Step~\ref{step:vboxes-init1} we initialized $ X_{\0} $ with $ 0 $ while at Step~\ref{step:vboxes-insert} we set $ X_{\0} = 1 $ since $ v_1 = 1 $. Hence, we execute the following procedure, described in Algorithm~\ref{alg:zero-vrect-dec}.
\begin{algorithm}
	\caption{Zero $ V $-Boxes Free Decoding}\label{alg:zero-vrect-dec}
	\begin{algorithmic}[1]
		\While{$X_{\0} = 1$}
		\State{Extract $ \bfu,A $ from the ($ d\log_q(n) + \log_q(f_d(V)) + 1 $)-prefix of $ SD(X) $}
		\State{Remove $ |A| $ entries from the start of $ X $}
		\State{Insert zero rectangle $ X_{\bfu + A}  = \0$ }
		\EndWhile
		\State{Return $ W = X_{[n]^d\setminus\{ \0 \}} $}
	\end{algorithmic}
\end{algorithm}

\subsection{The Boxes Unique Constraint} 
First, we use a union bound argument to derive a lower bound for $ V $ such that the redundancy of the set of $ V $-boxes unique arrays over $ \Sigma_q^{[n]^d} $ is at most $ 1 $.

\begin{lemma}\label{lem:vboxes-unique-red}
For $ V = 2d \log_q(n) + \frac{d-1}{d}\log_q(\log_q(n))+ \cO(1) $, and for $ n $ large enough it holds that $ |\cD\cA_{d,q}(n,V)| \ge q^{n^d-1} $. That is, $ \ered{\cD\cA_{d,q}(n,V)} \le 1$.
\end{lemma} 
\begin{IEEEproof}
	Let $ V = 2d\log_q(n) + \frac{d-1}{d} \log_q(\log_q(n)) + C + \log_q(\frac{q}{q-1}) $ for a positive constant $ C $ that will be determined later. 
	If an array $ W \in \Sigma^{[n]^d} $ is not $ V $-boxes unique, then it contains at least two identical boxes with a coordinates set $ A \in F_d(V) $. Hence, according to the union bound, the number of arrays that are not $ V $-boxes unique can be bounded from above by
		\begin{equation*}\label{eq:vboxes-unique-red2}
	\begin{aligned}	
	n^{2d} C_d V^{\frac{d-1}{d}} &\cdot q^{n^d-V} = q^{n^d} \cdot \frac{n^{2d} C_d V^{\frac{d-1}{d}}}{q^{V}}
	\\&= (q-1)q^{n^d-1} \cdot \frac{C_d V^{\frac{d-1}{d}}}{q^{\frac{d-1}{d}\log_q(\log_q(n)) + C}}
	\\&\overset{(a)}{\le}  (q-1)q^{n^d-1} \cdot \frac{C_d( (d+1)\log_q(n))^{\frac{d-1}{d}}}{\log_q(n)^{\frac{d-1}{d}}q^C}
	\\&\overset{(b)}{\le} (q-1)q^{n^d-1},
	\end{aligned}
	\end{equation*}
	where inequality (a) follows from $ V \le (d+1)\log_q(n) $ for $n$ large enough and (b) holds for $ n $ large enough by choosing a constant $ C \ge \log_q(C_d (d+1)^{\frac{d-1}{d}}) $.
	This accordingly implies that $ |\cD\cA_{d,q}(n,V)| \ge q^{n^d-1} $.
\end{IEEEproof}

When comparing the result of Lemma~\ref{lem:vboxes-unique-red} with the lower bound derived in Theorem~\ref{lem:lcubes-unique-red} for arrays that are $ L $-cubes unique, it follows that for the same volume $ V = L^d $, the minimal volume required for a redundancy of one symbol in the latter case is smaller by $ \Delta = \frac{d-1}{d}\log_q(\log_q(n)) + \cO(1) $. Note that this result of $\Delta$ is the same as the one achieved for the comparison of the zero-free constraints.

Next we find a lower bound on the value of $V$ which guarantees that the asymptotic rate of $\cD\cA_{d,q}(n,V)$ approaches 1. This is done similarly to the proof of Theorem 8 in \cite{EliGabMedYaa19IEEE}. The size of $ \cD\cA_{q,d}(n,V) $ will be estimated using a probabilistic approach. Consider the uniform distribution over all length-$ n $ sequences, then
$$ |\cD\cA_{q,d}(n,V)| = q^{n^d} \cdot Pr(W \in \cD\cA_{q,d}(n,V)).  $$
The asymptotic rate of $\cD\cA_{q,d}(n,V)$ is given by
\begin{equation}\label{eq:da-rate}
\begin{aligned}
\mathbb{R}_{q,d}(V) &\triangleq \limup{n}\frac{\log_q(|\cD\cA_{q,d}(n,V)|)}{n^d} 
\\&= 1 + \limup{n}\frac{1}{n^d}\log_q(Pr(W \in \cD\cA_{q,d}(n,V))).
\end{aligned}
\end{equation}

\begin{theorem}\label{th:da-rate}
Let $ n $ be an integer. For fixed $ d $, and $$ V = ad \log_q(n), $$ with $ a > 1 $, the asymptotic rate of $\cD\cA_{q,d}(n,V)$ approaches $ 1 $.
\end{theorem}

We prove Theorem~\ref{th:da-rate} using the asymmetric Lo\`asz local lemma which was first proved in~\cite{ErLo73} and is stated next as it appears in~\cite{AlSp2000}. 
\begin{lemma}[\cite{AlSp2000}, Lemma~5.1.1]~\label{lem:lll}
Let $ Y_0, \dots, Y_{m-1} $ be events in the arbitrary probability space. Let $ G = (V,E) $ be a graph with $ V = [m] $ such that for every $ i \in [m] $, the event $ Y_i $ is mutually independent of all the events $ \{Y_j \mid (i,j) \not \in E \} $. Suppose  that there are real numbers $ \alpha_0, \dots, \alpha_{m-1} $ such that $ \alpha_i \in [0,1] $ and for all $ i \in [m] $, $$ Pr(Y_i) \le \alpha_i \prod_{(i,j)\in E}(1-\alpha_j). $$ 
Then, it is satisfied that $$ Pr \left(  \bigcap_{i \in [m]} \overline{Y}_{i}\right) \ge \prod_{i \in [m]}(1-\alpha_i) $$
where $\overline{Y}_{i}$ is the complement of $ Y_i $.
\end{lemma}

\begin{IEEEproof}[Proof of Theorem~\ref{th:da-rate}]
Let $ X \in \Sigma_q^{[n]^d} $ be a random array in which each coordinate is chosen uniformly and independently over $ \Sigma_q $.
For coordinates $ \bfu, \bfv \in [n]^d $ and a set $ A \in F_d(V) $ such  that $ \bfu + A \subseteq [n]^d $ and $ \bfv + A \subseteq [n]^d $, we notate $ \bfz = (\bfu, \bfv, A) $ and denote  $ I_{\bfz} = \1(X_{\bfv + A} =X_{\bfu + A}) $, the indicator function of the event that the $ V $-boxes with coordinates set $ A $ that start at positions $ \bfu $ and $\bfv $ are identical. Let $$\hspace{-.25ex} \cZ \hspace{-.25ex}=\hspace{-.25ex} \{(\bfu, \bfv, A) \hspace{-.25ex}\mid\hspace{-.25ex} \bfu \neq \bfv, A \hspace{-.25ex}\in F_d(V), \bfu \hspace{-.25ex}+\hspace{-.25ex} A\hspace{-.25ex}\subseteq\hspace{-.25ex} [n]^d, \bfv\hspace{-.25ex} +\hspace{-.25ex} A\hspace{-.25ex} \subseteq [n]^d\} $$ be the set of all admissible triples, and notice that we are interested in a lower bound on 
$$ Pr(W \in \cD\cA_{q,d}(n,V)) = Pr \left(\sum_{\bfz \in \cZ} I_{\bfz} = 0\right). $$

Note that for every $ \bfz \in \cZ $ it holds that $Pr(I_\bfz) = \frac{1}{q^V} $. Let $ \bfz_0 = (\bfu_0, \bfv_0, A_0), \bfz_1 = (\bfu_1, \bfv_1, A_1) \in \cZ  $. It is clear that if the $ V $-boxes $ \bfu_0 + A_0 ,\bfv_0 + A_0 $ do not overlap with $ \bfu_1 + A_1 $ or $\bfv_1 + A_1$, then the indicators $ I_{\bfz_0}, I_{\bfz_1} $ are independent. We use Lemma~\ref{lem:lll} with a graph $ G = (V,E) $ such that $ V =  \cZ $ and there is an edge $ \bfz_0 \rightarrow \bfz_1 $  if at least one of $ \bfu_0 + A_0 ,\bfv_0 + A_0 $ overlaps with $ \bfu_1 + A_1$ or $\bfv_1 + A_1$. Thus, every $ \bfz  $

For a given minimal $ V $-box with coordinates set $ A_0 = [x_0] \times \cdots \times [x_{d-1}] $ at position $ \bfu $, an intersecting minimal $ V $-box with coordinates set $ A_1 = [y_0] \times \cdots \times [y_{d-1}] $ can only start at position that belongs to $$ U = \otimes_{i=0}^{d-1} [u_i -y_i+1, u_i + x_i-1].$$ 
The size of $ U $ can be bounded from above by
$$ |U| = \prod_{i=0}^{d-1} x_i +y_i -1 \le \prod_{i=0}^{d-1} x_iy_i \le A_0 \cdot A_1 \leq 4V^2,$$
where the last inequality holds since the size of every minimal $V$-box is bounded from above by $2V$. 
Therefore, since the number of minimal $ V $-boxes is at most $ f_d(V) \le c_d V^{\frac{d-1}{d}} $, the number of neighbors of each vertex is bounded from above by $$ 2 \cdot 4V^2 \cdot f_d(V) \cdot n^d \le c_d 8 V^{\frac{3d-1}{d}} n^d = c'_d V^{\frac{3d-1}{d}} n^d, $$
where $ c'_d = 8c_d $.

We set the numbers $ \alpha_\bfz = \frac{1}{c'_d V^{\frac{3d-1}{d}} n^d} $ for every $ \bfz \in \cZ $. It holds that
$$ \prod_{(\bfz, \bfz_1) \in E} (1-\alpha_{\bfz_1})  \geq \left(1 - \frac{1}{c'_d V^{\frac{3d-1}{d}} n^d} \right)^{c'_d V^{\frac{3d-1}{d}} n^d} \ge \frac{1}{e}$$
for every $ \bfz \in \cZ $ since the last expression approaches $ e^{-1} $ from above as {$ n~ \rightarrow~\infty $}. Hence, 
 the condition of the lemma holds since for every $ \bfz \in \cZ $,
 $$ Pr(I_\bfz) = \frac{1}{q^V} = \frac{1}{n^{ad}} \le \frac{1}{c'_d V^{\frac{3d-1}{d}} n^d} \cdot \frac{1}{e} \le \alpha_\bfz \prod_{(\bfz, \bfz_1) \in E} (1-\alpha_{\bfz_1}), $$
 where the first inequality holds since $c'_d V^{\frac{3d-1}{d}} n^d = o(n^{ad})$. By applying Lemma~\ref{lem:lll} we obtain
\begin{align*}
	Pr(W \in \cD\cA_{q,d}(n,V)) &\ge \prod_{\bfz \in \cZ}\left(1 - \frac{1}{c'_d V^{\frac{3d-1}{d}} n^d} \right)
	\\&\ge \left(1 - \frac{1}{c'_d V^{\frac{3d-1}{d}} n^d} \right)^{c_d V^{\frac{d-1}{d}} n^{2d}},
\end{align*}
since $ 1 - \frac{1}{c'_d V^{\frac{3d-1}{d}} n^d}\le 1 $ and $ |\cZ| \le f_d(V) n^{2d} \leq c_d V^{\frac{d-1}{d}}n^{2d}$.

Moreover, since 
$$ \left(1 - \frac{1}{c'_d V^{\frac{3d-1}{d}} n^d} \right)^{c_d V^{\frac{d-1}{d}} n^{2d}} \approx \exp \left(-\frac{n^d}{8 (ad \log_q(n))^2} \right), $$ it follows that 
$	\frac{1}{n^{d}} \log_q (Pr(W \in \cD\cA_{q,d}(n,V))) $ approaches 0 
as $ n \rightarrow \infty $. By plugging into (\ref{eq:da-rate}) we conclude that $ {\mathbb{R}_{q,d}(V) = 1} $.
\end{IEEEproof}

\section{Redundancy Analysis for the Two-Dimensional Zero-Free Constraints}\label{sec:red}
In this section we revisit the zero free $ L $-cubes constraint and the zero free $ V $-boxes constraint for the two-dimensional case. We analyze the redundancy of the set of arrays satisfying these constraints and present lower and upper bounds on the redundancy for both constraints. These bounds give an expression that is  asymptotically tight for the redundancy of $\cC_{2,q}(n,L)$ when $  n - 2L = \Theta(n)$ and the redundancy of $\cC\cA_{2,q}(n,L) $ when $ n - 2\sqrt{V} = \Theta(n)$.

\subsection{The Redundancy of the Zero $ L $-Squares Free Constraint}\label{subsec:red-c2qnl}
The result of this section is summarized in the following theorem.
\begin{theorem}\label{th:c2qnl}
There exist constants $ C_1,C_2 $ such that for any positive integer $ n $ it holds that 
$$ C_2 \frac{(n-2L)^2}{q^{L^2}} \le  \text{red}(\cC_{2,q}(n,L))\le  C_1 \frac{n^2}{q^{L^2}}.$$
\end{theorem}
The proof of Theorem~\ref{th:c2qnl} is given by lower and upper bounds proved in Claim~\ref{clm:c2qnl-lower} and Claim~\ref{clm:c2qnl-upper}, respectively. The next corollary follows immediately.
\begin{corollary}
	Let $ n,L $ be integers such that $ n - 2L = \Theta(n) $. Then, $$ \text{red}(\cC_{2,q}(n,L)) = \Theta\left(\frac{n^2}{q^{L^2}}\right). $$
\end{corollary}

An upper bound on the redundancy of $\cC\cA_{2,q}(n,L) $ is proved in the next claim.
\begin{claim}\label{clm:c2qnl-lower}
There exists a constant $ C_1 $ such that for any integer~$ n $ it holds that 
$$ \text{red}(\cC_{2,q}(n,L))\le  C_1 \frac{n^2}{q^{L^2}}.$$
\end{claim}
\begin{IEEEproof}
Let $ k $ be an integer, and let $ A_q(k,L) $ denote a set of squares over $ \Sigma^{[k]^2}_q $ that contain a zero $ (L/2) $-square in one of its corners, or a zero $ (L,L/2) $-rectangle at its right or left side, or a zero $ (L/2,L) $-rectangle at its upper or bottom side. That is,
\begin{align*}
	&A_q(k,L) \\ &\hspace{-.5ex}=\hspace{-.5ex}\left\{ X \text{ }  \middle\vert \hspace{-.5ex}
	\begin{array}{l}
		\exists i,j \in \{0,k-\frac{L}{2}\} \text{ s.t. } X_{(i,j)+[\frac{L}{2}]^2} = \0 \text{ or} \\
		\exists i \in \{0,k-\frac{L}{2}\}, j \in [k\hspace{-.5ex}-\hspace{-.5ex}L\hspace{-.5ex}+\hspace{-.5ex}1] \text{ s.t. }	X_{(i,j)+[\frac{L}{2}]\times[L]} = \0 \text{ or}\\
		\exists i \in [k\hspace{-.5ex}-\hspace{-.5ex}L\hspace{-.5ex}+\hspace{-.5ex}1], j \in \{0,k-\frac{L}{2}\} \text{ s.t. }	X_{(i,j)+[L]\times[\frac{L}{2}]} = \0.
	\end{array}\hspace{-1ex}\right\}.
\end{align*}
Note that $$ |A_q(k,L)| \le 4kq^{k^2-\frac{L^2}{2}} + 4 q^{k^2-\frac{L^2}{4}}. $$
Next, let $ B_q(k,L) = \cC_{2,q}(k,L) \setminus A_q(k,L) $. From Lemma~\ref{lem:lcubes-unique-red} we know that for $ L \ge \sqrt{2\log_q(k) + \log_q(\frac{q}{q-1})} $, then $|\cC_{2,q}(k,L)| \ge q^{k^2-1} $. This applies that for $ k\le q^{\frac{L^2-\log_q(\frac{q}{q-1})}{2}} $ we have
\begin{equation}\label{eq:11}
	\begin{aligned}
 |B_q(k,L)| &\ge q^{k^2-1} - 4kq^{k^2-\frac{L^2}{2}} - 4 q^{k^2-\frac{L^2}{4}}
 \\&\ge  q^{k^2-1} \left(1 - \frac{4k}{q^{\frac{L^2}{2}-1}} - \frac{4}{q^{\frac{L^2}{4}-1}}  \right).
 	\end{aligned}
\end{equation} 
We choose $  k = q^{\frac{L^2-7}{2}} $ which satisfies (\ref{eq:11}), and assume w.l.o.g that $ n \bmod k = 0 $.
Let $ E_q(n,L) $ denote the set of $ n $-squares that are composed of a grid of $ n^2/k^2 $ squares from $ B_q(k,L) $. We prove next that $ E_q(n,L) \subseteq \cC_{2,q}(n,L) $. Assume otherwise that $ X \in E_q(n,L) $ contains a corner zero $ L $-square. It is clear that the zero square is not contained in one of the $ B_q(k,L) $ $ k $-squares, and therefore it intersects two or four $ k $-squares. If it intersects two $ k $-squares, one of them must contain a zero $ (L,L/2) $-rectangle at its right of left edge or a zero $ (L/2,L) $-rectangle at its upper or bottom edge, which is a contradiction. Otherwise, the zero $ L $-square intersects four $ k $-squares and hence one of them contains a zero $ (L/2) $-square which contradicts the assumption as well. 
Thus, 
\begin{align*}
|&\cC_{2,q}(n,L)| \ge |B_q(q^{\frac{L^2-7}{2}},L)|^{\frac{n^2}{q^{L^2-7}}}
\\&= \left( q^{q^{L^2-7}-1} \left(1 - \frac{4q^{\frac{L^2-7}{2}}}{q^{\frac{L^2}{2}-1}} - \frac{4}{q^{\frac{L^2}{4}-1}} \right) \right)^{\frac{n^2}{q^{L^2-7}}}
\\&= q^{n^2} \cdot q^{-\frac{n^2}{q^{L^2-7}}} \left(1 - \frac{4}{q^{2.5}} - \frac{4}{q^{\frac{L^2}{4}-1}} \right)^{\frac{n^2}{q^{L^2-7}}}
\\&= q^{n^2}  (q^{-1}(1 - \frac{4}{q^{2.5}}))^{\frac{n^2}{q^{L^2-7}}} \left(1 -  \frac{4}{q^{\frac{L^2}{4}-1}(1-4q^{-2.5})} \right)^{\frac{n^2}{q^{L^2-7}}}.
\end{align*}
It is known that for all $ x < -1 $, $ (1+\frac{1}{x})^{x+1} < e $. We denote $ x = - \frac{q^{\frac{L^2}{4}-1}(1-4q^{-2.5})}{4} $. For $ L \ge 3 $ and $ q \ge 2 $ we have that $ x < -1  $ and hence 
\begin{align*}
\left(1 -  \frac{4}{q^{\frac{L^2}{4}-1}(1-4q^{-2.5})} \right)^{\frac{n^2}{q^{L^2-7}}}
&=  (1+\frac{1}{x})^{(x+1)(\frac{n^2}{q^{L^2-7}})/(x+1)}
\\&\overset{(a)}{>} \exp \left( (\frac{n^2}{q^{L^2-7}})/(x+1)\right)
\\&\overset{(b)}{=} \exp \left( (c_2\frac{n^2}{q^{L^2}})/(-c_1 q^{\frac{L^2}{4}})\right)
\\&= \exp\left({-\frac{c_2}{c_1} \frac{ n^2}{q^{\frac{5L^2}{4}} }}\right)
\end{align*}
where (a) follows from $ x + 1 < 0 $ and (b) follows from a choice of appropriate constants $ c_1,c_2 $. Finally, let $ c_3 = \frac{1}{1-4q^{-2.5}} $ for some constant $ c_3 > 0 $. We conclude that 
$$ |\cC_{2,q}(n,L)| \ge q^{n^2}\cdot (qc_3)^{-c_2\frac{n^2}{q^{L^2}}}\exp({-\frac{c_2}{c_1} \frac{ n^2}{q^{\frac{5L^2}{4}} }})$$
and thus
$$ \text{red}(\cC_{2,q}(n,L)) \le c_2(1 + \log_q(c_3))\frac{n^2}{q^{L^2}} + \log_q(e)\frac{c_2}{c_1} \frac{ n^2}{ q^{\frac{5L^2}{4}} }.  $$ 
It follows that there exists a constant $ C_1 > 0 $ such that 
\begin{equation}\label{eq:comp-red-c2nl}
\text{red}(\cC_{2,q}(n,L))\le  C_1 \frac{n^2}{q^{L^2}}. 
\end{equation}

Note that before constructing $ E_q(n,k) $, if $ n \bmod k \neq 0 $ we can pick $ n' = n + (n - ( n \bmod k)) $ and continue the proof for $ n' $ instead of $ n $ to receive the result of (\ref{eq:comp-red-c2nl}) for $ n' $. Since $ n' - n \le c_2 q^{\frac{L^2}{2}} $, this affects only the constant $ C_1 $ and the claim statement holds for $ n $.
\end{IEEEproof}

Next, a lower bound on $\text{red}(\cC_{2,q}(n,L))$ is given.
\begin{claim}\label{clm:c2qnl-upper}
	There exists a constant $ C_2 $ such that for any integer~$ n $ it holds that  $$ \text{red}(\cC_{2,q}(n,L)) \ge C_2 \frac{(n-2L)^2}{q^{L^2}} $$
\end{claim}
\begin{IEEEproof}
Let $ D_q(n,L) $ denote the set of $ n $-squares that are constructed from a grid of $ \cC_{2,q}(2L,L) $ squares, i.e., $ 2L $-squares that are zero $ L $-square free. The remained $ n^2 - (\floorenv{\frac{n}{2L}}L)^2 $ entries are filled with any symbols from $ \Sigma_q $. We have that  $ \cC_{2,q}(n,L) \subseteq  D_q(n,L) $ and hence,
\begin{align*}
|\cC_{2,q}(n,L)| \le  |\cC_{2,q}(2L,L)|^{(\floorenv{\frac{n}{2L}})^2}\cdot q^{n^2 - (\floorenv{\frac{n}{2L}}L)^2}.
\end{align*}
Let $ \beta(L) $ denote the set of $ 2L $-squares that contain a zero $ L $-square exactly once. We lower bound $ |\beta(L)| $ by placing a zero $ L $-square at some position $ (i,j) $ for $ X \in \beta(L) $, and adding redundancy symbols to ensure that no other zero $ L $-squares exist in $ X $. Assume w.l.o.g that $ i,j \in [1,L-1] $, i.e., the zero square is in the middle of $ X $. We set four non-zero symbols at positions $ I = \{(i,j-1),(i+L,j),(i+L-1,j+L), (i-1, j+L-1) \} $. For example, let $ L = 3, i=j=1 $, and
$$ X = 
\begin{pmatrix}
	 &  &  & 1  &  &  \\
	1 & 0 & 0 & 0  &  &  \\
	 & 0 & 0 & 0  &  &  \\
	 & 0 & 0 & 0  & 1 &  \\
	 & 1 &  &   &  &  \\
	 &  &  &   &  &  \\
\end{pmatrix} 
$$ has a exactly one zero $ L $-square. 
Note that besides $ X_{(i,j)+L^2} $, every $ L $-square in $ X $ contains one of the coordinates of $ I $ and hence $ X_{(i,j)+L^2} $ is the only zero $ L $-square in $ X $. If the zero square is next to one of the sides of $ X $, it is enough to set only the valid positions of $ I $ in order to eliminate additional zero $ L $-squares.
 Hence, $$ |\beta(L)| \ge (L+1)^2(q-1)^4q^{3L^2-4}. $$
Since $ \beta(L) \cap \cC_{2,q}(2L,L) = \emptyset $  we can write
\begin{align*}
|\cC_{2,q}(2L,L)| &\le q^{4L^2} - L^2(q-1)^4q^{3L^2-4} 
\\&= q^{4L^2} \left(1 - \frac{L^2(q-1)^4}{q^{L^2+4}}  \right)
\end{align*}
and by combining the inequalities we have 
\begin{align*}
| \cC_{2,q}(n,L) | &\le \left( q^{4L^2} \left(1 - \frac{L^2(q-1)^4}{q^{L^2+4}}  \right) \right)^{(\floorenv{\frac{n}{2L}})^2} \cdot q^{n^2 - (\floorenv{\frac{n}{2L}}L)^2}
\\&=q^{n^2} \left(1 - \frac{L^2(q-1)^4}{q^{L^2+4}}  \right)^{(\floorenv{\frac{n}{2L}})^2}
\\&\le q^{n^2} \left(\exp(-\frac{L^2(q-1)^4}{q^{L^2+4}})\right)^{(\floorenv{\frac{n}{2L}})^2}
\\&\overset{(a)}{\le} q^{n^2 - \log_q(e)\frac{L^2(q-1)^4}{q^{L^2+4}}\cdot (\frac{n}{2L}-1)^2}
\\&\le q^{n^2 - \log_q(e)\frac{(q-1)^4(n-2L)^2}{4q^{L^2+4}}},
\end{align*}
where (a) follows from the inequality $ (1-x) < e^{-x} $ for all $ x $. By denoting $ C_2 = \frac{\log_q(e)(q-1)^4}{4q^4}  $ we can conclude that $$ \text{red}(\cC_{2,q}(n,L)) \ge C_2 \frac{(n-2L)^2}{q^{L^2}}. $$
\end{IEEEproof}

\subsection{The Redundancy of the Zero $V$-Boxes Free Constraint}
Next, we present tight bounds on the cardinality of $ \cC\cA_{2,q}(n,V) $. These bounds use similar methods to those presented in Section~\ref{subsec:red-c2qnl}. Nonetheless, we introduce improvements and adaptations to those methods in order to fit the constraint where the zero sub-arrays are rectangles and only their area is known. The main result is summarized in the next theorem.
\begin{theorem}\label{th:ca2qnl}
There exist constants $ C'_1,C'_2 $  such that for any integer $ n $ it holds that $$ C'_2 \frac{(n-2\sqrt{V})^2}{q^{V - \log_q(V)}} \le  \text{red}(\cC\cA_{2,q}(n,V))\le  C'_1 \frac{n^2}{q^{V - \log_q(V)}}.$$
\end{theorem}
The proof of Theorem~\ref{th:ca2qnl} is given by lower and upper bounds proved in Claim~\ref{clm:ca2qnl-lower} and Claim~\ref{clm:ca2qnl-upper}, respectively. The next corollary follows immediately.
\begin{corollary}
	Let $ n,V $ be integers such that $ {n - 2\sqrt{V} = \Theta(n)} $. Then, $$ \text{red}(\cC\cA_{2,q}(n,V)) = \Theta\left(\frac{n^2}{q^{V - \log_q(V)}}\right). $$
\end{corollary}

\begin{claim}\label{clm:ca2qnl-lower}
	There exists a constant $ C'_1 $ such that for any integer~$ n $ it holds that $$ \text{red}(\cC\cA_{2,q}(n,V))\le  C'_1 \frac{n^2}{q^{V - \log_q(V)}}.$$
\end{claim}
\begin{IEEEproof}
Let $ k $ be an integer, and let $ A_q(k,V) $ denote a set of squares over $ \Sigma^{[k]^2}_q $ that contain a zero $ (V/4) $-rectangle at one of its corners, or a zero $ (V/2) $-rectangle at any of its sides. The following upper bound holds for the size of the set $A_q(k,V)$.
\begin{align*}
|A_q(k,V)| &\le  4k f_2(\frac{V}{2}) q^{k^2 - \frac{V}{2}} + 4 f_2(\frac{V}{4}) q^{k^2-\frac{V}{4}}
\\& \le  4k\sqrt{2V} q^{k^2 - \frac{V}{2}} + 4\sqrt{V}q^{k^2-\frac{V}{4}}.
\end{align*}
Let $ B_q(k,V) = \cC\cA_{2,q}(k,V) \setminus  A_q(k,V) $. From Lemma~\ref{lem:vboxes-free-red2}, if $ k  $ satisfies
\begin{equation}\label{eq:upper-ca}
	\begin{aligned}
	k \le q^{\frac{1}{2}(V - \frac{1}{2}\log_q(V) + \log_q(\frac{q-1}{2q}))},
	\end{aligned}
\end{equation}
then $ \cC\cA_{2,q}(k,V) \ge q^{k^2-1} $ and therefore, 
\begin{align*}
|B_q(k,V)| &\ge q^{k^2-1} - 4k\sqrt{2V} q^{k^2 - \frac{V}{2}} - 4\sqrt{V} q^{k^2-\frac{V}{4}} 
\\&\ge q^{k^2-1}\left( 1 - \frac{4k\sqrt{2V}}{q^{\frac{V}{2}-1}} - \frac{4\sqrt{V}}{q^{\frac{V}{4}-1}}  \right).
\end{align*}
We pick $ k = q^{\frac{V}{2}-\frac{1}{2}\log_q(V)-4} $ which satisfies equation (\ref{eq:upper-ca}), and assume w.l.o.g that $ n \bmod k = 0 $. 

Next, we construct $ E_q(n,V) $, which is the set of $ n $-squares that are composed of a grid of $ n^2/k^2 $ squares from $ B_q(k,V) $. It can be shown similarly to the proof of Claim~\ref{clm:c2qnl-lower} that 
$ E_q(n,V) \subseteq \cC_{2,q}(n,V) $ and thus,
\begin{align*}
&|\cC\cA_{2,q}(n,V)| \ge |B_q(k,V)|^{\frac{n^2}{k^2}}
\\&=
 |B_q(q^{\frac{V}{2}-\frac{1}{2}\log_q(V)-4},V)|^{\frac{n^2}{q^{V-\log_q(V)-8}}}
\\&\ge\hspace{-0.5ex} \left(q^{q^{V-\log_q(V)-8}-1} \left( 1\hspace{-0.5ex}-\hspace{-0.5ex}\frac{4\sqrt{2V} q^{\frac{V}{2}-\frac{1}{2}\log_q(V)-4}}{q^{\frac{V}{2}-1}}\hspace{-0.5ex}-\hspace{-0.5ex}\frac{4\sqrt{V}}{q^{\frac{V}{4}-1}}  \right) \right)^{\frac{Vn^2}{q^{V-8}}}
\\&= q^{n^2}q^{-\frac{Vn^2}{q^{V-8}}}\left( 1 - \frac{4\sqrt{2}}{q^3} - \frac{\sqrt{V}}{q^{\frac{V}{4}-1}}  \right)^{\frac{Vn^2}{q^{V-8}}}
\\&= q^{n^2}(q^{-1}(1 - \frac{4\sqrt{2}}{q^3}))^{\frac{Vn^2}{q^{V-8}}}\left( 1 - \frac{\sqrt{V}}{q^{\frac{V}{4}-1}(1 - \frac{4\sqrt{2}}{q^3})}  \right)^{\frac{Vn^2}{q^{V-8}}}.
\end{align*}
It is known that for all $ x < -1 $, $ (1+\frac{1}{x})^{x+1} < e $. We denote $$ x = - \frac{q^{\frac{V}{4}-1}(1 - \frac{4\sqrt{2}}{q^3})}{\sqrt{V}}. $$ For $ V \ge 2^5 $ and $ q \ge 2 $ we have that $ x < -1  $ and hence 
\begin{align*}
&\left( 1 - \frac{\sqrt{V}}{q^{\frac{V}{4}-1}(1 - \frac{4\sqrt{2}}{q^3})}  \right)^{\frac{Vn^2}{q^{V-8}}}
\\&=  (1+\frac{1}{x})^{(x+1)(\frac{Vn^2}{q^{V-8}})/(x+1)}
\\&\overset{(a)}{>} \exp\left((\frac{n^2}{q^{V-\log_q(V)-8}})/(x+1)\right)
\\&\overset{(b)}{=} \exp\left((c_2\frac{n^2}{q^{V-\log_q(V)}})/(-c_1 q^{\frac{V}{4}-\frac{1}{2}\log_q(V)})\right)
\\&= \exp\left({-\frac{c_2}{c_1} \frac{ n^2}{q^{\frac{5V}{4}-\frac{3}{2}\log_q(V)}}}\right),
\end{align*}
where (a) follows from $ x + 1 < 0 $ and (b) follows from a choice of appropriate constants $ c_1,c_2 > 0 $. Finally, we denote $ c_3 = (1 - \frac{4\sqrt{2}}{q^3})^{-1} $ and conclude that 
$$ |\cC\cA_{2,q}(n,L)| \ge q^{n^2} (qc_3)^{-c_2\frac{n^2}{q^{V-\log_q(V)}}}\exp\left({-\frac{c_2}{c_1} \frac{ n^2}{q^{\frac{5V}{4}-\frac{3}{2}\log_q(V)}}}\hspace{-0.5ex}\right)$$
and thus the redundancy satisfies
\begin{align*}
\text{red}(\cC\cA_{2,q}(n,V)) &\le c_2(1 + \log_q(c_3))\frac{n^2}{q^{V-\log_q(V)}} 
\\&+ \log_q(e)\frac{c_2}{c_1} \frac{ n^2}{q^{\frac{5V}{4}-\frac{3}{2}\log_q(V)}}.
\end{align*}
It follows that there exists a constant $ C'_1 > 0 $ such that 
$$ \text{red}(\cC\cA_{2,q}(n,V)) \le C'_1 \frac{n^2}{q^{V-\log_q(V)}}.  $$ 
Similarly to the proof of Claim~\ref{clm:c2qnl-lower}, when $ n \bmod k \neq 0 $ we can enlarge $ n $ to the closest multiple of $ k $, and the claim statement still holds for $ n $.
\end{IEEEproof}

\begin{claim}\label{clm:ca2qnl-upper}
	There exists a constant $ C'_2 $ such that for any integer~$ n $ it holds that $$ \text{red}(\cC\cA_{2,q}(n,V)) \ge C'_2 \frac{(n-2\sqrt{V})^2}{q^{V - \log_q(V)}}. $$
\end{claim}
\begin{IEEEproof}
Let $ D_q(n,V) $ denote the set of $ n $-squares that are constructed from a grid of $ \cC\cA_{2,q}(2\sqrt{V},V) $ squares, i.e., $ 2\sqrt{V} $-squares that are $ V $-rectangles free, where the remained entries are filled with any symbols from $ \Sigma_q $. We have that  $ \cC\cA_{2,q}(n,V) \subseteq  D_q(n,V) $ and therefore,
\begin{align}\label{eq:upper-ca1}
|\cC\cA_{2,q}(n,V)| \le  |\cC\cA_{2,q}(2\sqrt{V},V)|^{(\floorenv{\frac{n}{2\sqrt{V}}})^2}q^{n^2 - (\floorenv{\frac{n}{2\sqrt{V}}})^2 4V}
\end{align}
Let $ \beta(V) $ denote the set of ($2\sqrt{V}$)-squares that contain a exactly one zero $ V $-rectangle. 
Similarly to Claim~\ref{clm:c2qnl-upper}, we lower bound $ \beta(V) $ by placing a zero $ V $-rectangle in some $ X \in \beta(V) $ and using redundancy symbols to ensure that no other zero $ V $-rectangles exist in $ X $. Let $ X_{(i,j)+A} $ be such a zero $ V $-rectangle, and since $ X $ is a ($2\sqrt{V}$)-square, the shorter side of $ A $ is in the range $ [\sqrt{V}/2, \sqrt{V}] $ and therefore there are $ \sqrt{V} $ possible options for $ A $. Moreover, there are at most $ (\sqrt{V}+1)^2 $ different possible options for the indexes $ i,j $. 

Next, in order to eliminate additional zero $ V $-rectangles, it is enough to ensure that no additional zero $ (\sqrt{V}/2) $-squares exist in $ X $.
Assume w.l.o.g that $ A = [a]\times[b] $ where $ a \ge \sqrt{V}/2 $, and that the zero square is in the middle of $ X $. By surrounding the zero square with non zero symbols at positions $ I_1 = \{(i,j-1),(i+a/2,j-1),(i+a,j),(i+a-1,j+b), (i+a/2-1,j+b)(i-1, j+b-1) \} $, we ensure that no zero $ (\sqrt{V}/2) $-squares intersect with $ X_{(i,j)+A} $. Additionally, in order to prevent zero $ (\sqrt{V}/2) $-squares in the rest of $ X $, we set a non-zero symbol every $ \sqrt{V}/2 $ entries, as long as those do not intersect with $ X_{(i,j)+A} $; that is, $ I_2 = ([1,3] \cdot (\sqrt{V}/2))^2 \setminus ((i,j)+A)$. 
This results in at most $ c_1 = |I_1| + |I_2| \le 15 $ non-zero symbols. Therefore, we can bound 
\begin{align*}
|\beta(V)| &\ge (\sqrt{V}+1)^2 \sqrt{V} q^{3V - c_1}(q-1)^{c_1}
\\& \ge V^{1.5}q^{3V - c_1}(q-1)^{c_1} .
\end{align*}
Next, we have that $ \beta(V) \cap \cC\cA_{2,q}(2\sqrt{V},V) = \emptyset $ and hence,
\begin{equation}\label{eq:upper-ca2}
\begin{aligned}
|\cC\cA_{2,q}(2\sqrt{V},V)| &\le q^{4V} - V^{1.5}q^{3V - c_1}(q-1)^{c_1}
\\&= q^{4V} \left( 1 - \frac{V^{1.5}(q-1)^{c_1}}{q^{V+c_1}} \right).
\end{aligned}
\end{equation}
By combining inequalities (\ref{eq:upper-ca1}) and (\ref{eq:upper-ca2}) we get that $ |\cC\cA_{2,q}(n,V)|$ is bounded from above by
\begin{align*}
&\left( q^{4V} \left( 1 - \frac{V^{1.5}(q-1)^{c_1}}{q^{V+c_1}} \right) \right)^{(\floorenv{\frac{n}{2\sqrt{V}}})^2}q^{n^2 - (\floorenv{\frac{n}{2\sqrt{V}}})^2 4V}
\\&= q^{n^2} \left( 1 - \frac{V^{1.5}(q-1)^{c_1}}{q^{V+c_1}} \right)^{(\floorenv{\frac{n}{2\sqrt{V}}})^2}
\\&\le q^{n^2} \exp\left(-\frac{V^{1.5}(q-1)^{c_1}}{q^{V+c_1}} \right)^{(\floorenv{\frac{n}{2\sqrt{V}}})^2}
\\&\le q^{n^2} \exp\left(\frac{V^{1.5}(q-1)^{c_1}}{q^{V+c_1}}(\frac{n}{2\sqrt{V}}-1)^2\right)
\\&= q^{n^2} \exp\left(\frac{\sqrt{V}(q-1)^{c_1}(n-2\sqrt{V})^2}{4q^{V+c_1}}\right)
\\&= q^{n^2 - \log_q(e)\frac{\sqrt{V}(q-1)^{c_1}(n-2\sqrt{V})^2}{4q^{V+c_1} }} 
\end{align*}
and by denoting $ C'_2 = \frac{\log_q(e)(q-1)^{c_1}}{4q^{c_1}}  $ we have $$ \text{red}(\cC\cA_{2,q}(n,V)) \ge C'_2 \frac{(n-2\sqrt{V})^2}{q^{V - \log_q(V)}}. $$
\end{IEEEproof}

	\section{Conclusion}\label{sec:concl}
	This paper studied two main families of constraints, referred as the zero $ L $-cube free constraint and the $ L $-cube unique constraint, for multidimensional arrays that impose conditions on the cubes contained in the array. The paper studied also the extensions of these constraints to the case where the conditions are imposed on sub-arrays that are multidimensional boxes and not necessarily cubes, where only their volume is given as a parameter. For the zero free constraints, we presented a lower bound on the size of the sub-array such that the redundancy of the constraint is at most a single symbol, an efficient encoding algorithm for any dimension that uses a single redundancy symbol and tight bounds on the cardinality of the constraints specifically for the two-dimensional case. As for the cube-unique and box-unique constraints, we presented a lower bound on the size of the sub-array such that the asymptotic rate of the set of valid arrays approaches $ 1 $, as well as conditions for the redundancy to be at most a single symbol. Additionally, we presented an encoder for the two-dimensional $ L $-square unique constraint that uses a single redundancy bit.

	

\bibliographystyle{IEEEtranS}
\bibliography{mybib}
	
\end{document}